\documentclass[notitlepage,superscriptaddress,showpacs,nobalancelastpage,twocolumn,aps,longbibliography,prl]{revtex4-2}
\usepackage[english]{babel}
\RequirePackage[T1]{fontenc}
\RequirePackage{times} 

\usepackage{siunitx} 
\usepackage[italicdiff]{physics}
\usepackage{amsfonts}
\usepackage{lipsum}
\usepackage{multirow}
\usepackage{subfigure}
\usepackage{amsmath}
\usepackage{amssymb}
\usepackage{graphicx,epstopdf}
\usepackage{xspace}
\usepackage{array}
\usepackage[hidelinks]{hyperref}
\usepackage[dvipsnames]{xcolor}
\usepackage{my_symbols}
\usepackage{CJK}
\usepackage{bm}
\usepackage{verbatim}
\usepackage{tikz}
\usepackage{comment}
\usepackage{marginnote}
\usepackage[textwidth=1.5cm]{todonotes}
\usetikzlibrary{calc,3d}
\usetikzlibrary{arrows}
\usepackage[normalem]{ulem}
\usepackage{soul}
\usepackage{dsfont}
\usepackage{accents}

\sethlcolor{green}

{\par}

\newcounter{mycomment}

\includecomment{comment} 

\begin{document}

\begin{CJK*}{UTF8}{gbsn} 
\title{The Connection between Spin Wave Polarization and Dissipation}
\author{Yutian Wang (王昱天)}
\affiliation{Department of Physics and State Key Laboratory of Surface Physics, Fudan University, Shanghai 200433, China}
\author{Jiongjie Wang}
\affiliation{Department of Physics and State Key Laboratory of Surface Physics, Fudan University, Shanghai 200433, China}
\author{Ruoban Ma}
\affiliation{Department of Physics and State Key Laboratory of Surface Physics, Fudan University, Shanghai 200433, China}
\author{Jiang Xiao (萧江)}
\email[Corresponding author:~]{xiaojiang@fudan.edu.cn}
\affiliation{Department of Physics and State Key Laboratory of Surface Physics, Fudan University, Shanghai 200433, China}
\affiliation{Institute for Nanoelectronics Devices and Quantum Computing, Fudan University, Shanghai 200433, China}
\affiliation{Shanghai Research Center for Quantum Sciences, Shanghai 201315, China}
\affiliation{Shanghai Branch, Hefei National Laboratory, Shanghai 201315, China}

\begin{abstract}
    This study establishes a fundamental connection between the dissipation and polarization of spin waves, which are often treated as independent phenomena. Through theoretical analysis and numerical validation, we demonstrate that within the linearized spin wave regime, a spin wave mode's dissipation rate, defined as the ratio of linewidth to the resonance frequency, exceeds Gilbert damping by a factor given by its spatially averaged polarization. This average is governed by a non-positive definite weight, whose magnitude depends on the magnon density of the local excitation, while its sign is dictated by the local polarization handedness. Remarkably, this universal connection applies across diverse magnetic interactions and textures, offering crucial insights into spin wave dynamics and dissipation.
\end{abstract}
\maketitle
\end{CJK*}

Spin waves, or magnons in quantized form, are collective excitations of magnetic moments in magnetic materials \cite{stancil_spin_2009}. 
The study of spin waves bridges fundamental physics with practical applications, from understanding magnetic phase transitions to developing novel spintronic devices \cite{chumak_roadmap_2021}.
Similar to sound waves transmitting energy through vibrations, spin waves carry magnetic energy and angular momentum without the transport of charge. This unique property has made them increasingly attractive for next-generation information processing technologies with reduced energy consumption and faster processing, particularly in the field of magnonics \cite{demokritov_magnonics:_2012, yu_magnetic_2021, yuan_quantum_2022, zare_rameshti_cavity_2022}.

One of the critical challenges in utilizing spin waves for practical applications is their dissipation, which limits their lifetime and propagation length. This dissipation is phenomenologically described by the Gilbert damping in the Landau-Lifsthiz-Gilbert equation \cite{gilbert_phenomenological_2004}. Yttrium iron garnet (YIG) stands out with the lowest damping constant as low as \(10^{-5}\), enabling spin wave propagation up to several millimeters \cite{maendlSpinWavesLarge2017,qinPropagatingSpinWaves2018,schmidtUltraThinFilms2020}.
The realistic value of the damping constant is influenced by many factors, including the material composition, temperature, and extrinsic geometry or interfaces \cite{azzawiMagneticDampingPhenomena2017}. For magnonic devices to be viable in information processing, it is essential to enhance spin wave lifetimes and propagation lengths. Consequently, understanding and controlling dissipation has become a central focus in spin wave research.

The vector nature of magnetic moment endows spin waves with a polarization degree of freedom,
akin to the polarization in sound and optical waves. The spin wave polarization characteristics differ significantly between ferromagnetic and antiferromagnetic spin waves. 
In ferromagnets, the broken time-reversal symmetry restricts spin waves to only right-handed circular or elliptical polarization \cite{gurevichMagnetizationOscillationsWaves1996}. In contrast, antiferromagnets support both right- and left-handed circular polarizations, as well as their linear combinations, leading to diverse linear and elliptical polarizations \cite{gomonay_spintronics_2014,lanAntiferromagneticDomainWall2017}. This polarization versatility presents promising opportunities for magnonic information processing based on spin wave polarization manipulation \cite{cheng_antiferromagnetic_2016,lan_antiferromagnetic_2017,yuPolarizationselectiveSpinWave2018a,yu_magnetic_2020,han_birefringence-like_2020}, offering advantages over traditional methods that rely on amplitude or phase \cite{kostylevSpinwaveLogicalGates2005,schneider_realization_2008,kruglyakMagnonics2010,khitunNonvolatileMagnonicLogic2011,mahmoud_introduction_2020}.

As two important properties of spin wave, dissipation and polarization are usually considered as distinct phenomena. However, this work reveals a fundamental connection between the two, termed the dissipation-polarization connection. Early in the seventies, Kambersky \etal examined the impact of elliptically polarized eigenmodes on ferromagnetic resonance (FMR) linewidth \cite{kambersky_spin-wave_1975},
and Puszkarski proposed that elliptical spin precession modifies resonance line intensity \cite{puszkarski_theory_1979}. 
And more recently, Rozsa \etal pointed out a link between polarization and linewidth in soft mode excitations in Skyrmions \cite{rozsa_effective_2018}. 
Our findings suggest that the connection between spin wave dissipation and polarization transcends the specific cases in the these studies, asserting that spin wave dissipation rate, which is closely related to the linewidth of magnetic resonances, is strictly determined by the averaged polarization. We also point out that this connection between spin wave dissipation and polarization is universal: it applies to both ferromagnetic and antiferromagnetic spin waves across various interactions, including anisotropy, exchange, dipolar interactions, and complex magnetic textures. By establishing the dissipation-polarization connection, we enhance the understanding of both spin wave dissipation and polarization, pointing the way for improving spin wave lifetime and propagation length in magnonic applications.


\emph{Dissipation-Polarization Connection - }
The magnetization dynamics is governed by the Landau-Lifshitz-Gilbert equation 
\cite{gilbert_phenomenological_2004}
\begin{equation}
    \label{eqn:LLG}
    \dot \mb = - \gamma \mb \times \bH_{\rm eff}+\alpha \mb\times\dot\mb,
\end{equation} 
where $\gamma$ is gyromagnetic ratio, and $\alpha$ is the Gilbert damping parameter characterizing the dissipation. $\bH_\text{eff} = - \delta F/\delta \mb$ is the effective magnetic field derived from the free energy $F$. 
The resonant and dissipative behavior of spin wave mode can be characterized by its complex frequency $\omega$, whose real and imaginary parts represent the resonance frequency and the broadening (linewidth), respectively. We define the magnetic dissipation rate as the ratio of the linewidth to the frequency:
\begin{equation}
    \label{eqn:diss}
    \beta 
    \equiv \frac{\mbox{linewidth}}{\mbox{peak frequency}} 
    = \frac{\Im{\omega}}{\abs{\Re{\omega}}}.
\end{equation}
Patton pointed out that the frequency-swept linewidths, not field-swept linewidths, are proportional to the dissipation rate \cite{patton_linewidth_1968}.
The dissipation rate also characterizes the number of precession cycles before the spin wave is damped out, or the fraction of energy dissipated per cycle.
In most common case, the damping rate equals to the Gilbert damping constant, \ie $\beta = \alpha$. In the more general case as discussed in this paper, they are not identical, but $\beta \ge \alpha$.

Because all free energy contributions under consideration are quadratic in $\mb$, for an eigen mode with damping neglected, the tip of the magnetization vector follows along an elliptical trajectory in the complex plane formed by $m_x$ and $m_y$. This elliptical trajectory can be decomposed as a superposition of right- and left-handed motion: 
$
    \psi(t) = m_x + i m_y = m^+ e^{i\omega t}+m^- e^{-i\omega t}, 
$
with $m^\pm$ the corresponding (complex) amplitudes.
The lengths of semi-axis along the two principal axes of the elliptical trajectory are $a = |m^+|+|m^-|$ and $b = |m^+|-|m^-|$, respectively. 
And the spin wave polarization can be quantified by the signed ellipticity
\begin{equation}
    \label{eqn:pol}
    \eta = \frac{|m^+|-|m^-|}{|m^+|+|m^-|} = \frac{b}{a}
    \in [-1,1],
\end{equation}
for which $\eta > 0$, $\eta < 0$, and $\eta = 0$ correspond to the right-handed, left-handed polarizations, and linearly polarized along the major axis respectively. 
We may also reparameterize the ellipticity $\eta$ with a complex parameter $r$ with $\eta = e^{-2r}$, for which real $r$ corresponds to right-handed polarization and $r$ with $\arg{r} = \pi/2$ corresponds to left-handed one.


The dissipation rate and the polarization are seemly disconnect concepts. However, for linearized spin wave excitations, we now establish a simple but universal connection between them for the eigenmodes (proof in Appendix A):
\begin{equation}
    \label{eqn:betar}
    \beta
    = \alpha \frac{\eta + \eta^{-1}}{2}
    = \alpha \cosh(2r).
\end{equation}
Because $\abs{\cosh(2r)}>1$, it is evident that the damping rate $\abs{\beta}$ always exceeds the intrinsic Gilbert damping constant $\alpha$ as long as the spin wave is not circular ($\abs{\eta} \neq 1$ or $\Re{r}\neq 0$). A simple message is that elliptically polarized spin wave dissipates faster than the circular one.

In the more general case where the Gilbert damping $\alpha_\bx$ or the polarization $r_\bx$ vary in space, the dissipation rate $\beta_\bx = \alpha_\bx \cosh(2r_\bx)$ is a function of position and is higher (lower) at locations with more elliptical (circular) polarizations. Consequently, the overall dissipation rate of the whole system is given by a weighted average (proof in Appendix A):
\begin{equation}
    \label{eqn:dpc}
    \beta 
    = \expval{\beta_\bx}
    \equiv
    \frac{\int \dd^3{\bx}~S_\bx \alpha_\bx \cosh(2r_\bx)}{\int \dd^3{\bx}~S_\bx}.
\end{equation}
Since $b$ might take either sign, the weight $S_\bx = \pi ab 
= (\pi/\omega)\Im{\dot{\psi}(t)\psi^*(t)}$ here
is the directed (signed) area of enclosed by the local magnetization precession trajectory at position $\bx$, and the sign is determined by precession direction ($+$ for right-handed  and $-$ for left-handed). 
The dissipation-polarization connection in \Eq{eqn:dpc} is exact for linearized spin wave, \ie the full Hamiltonian is of quadratic form of local magnon creation and annihilation operators, regardless of the types of interactions included.
A similar expression to \Eq{eqn:dpc} has been proposed by Rozsa et al. \cite{rozsa_effective_2018} to explain the damping enhancement in non-collinear spin configurations, particularly in ferromagnetic Skyrmions. 
However, the formulation presented in \Eq{eqn:dpc}, which incorporates negative weights for the first time, emerges as a general principle that links dissipation and polarization. This connection is applicable across diverse scenarios, including those where both right-handed and left-handed excitations coexist. 

\begin{figure*}[ht]
    \includegraphics[width=\textwidth]{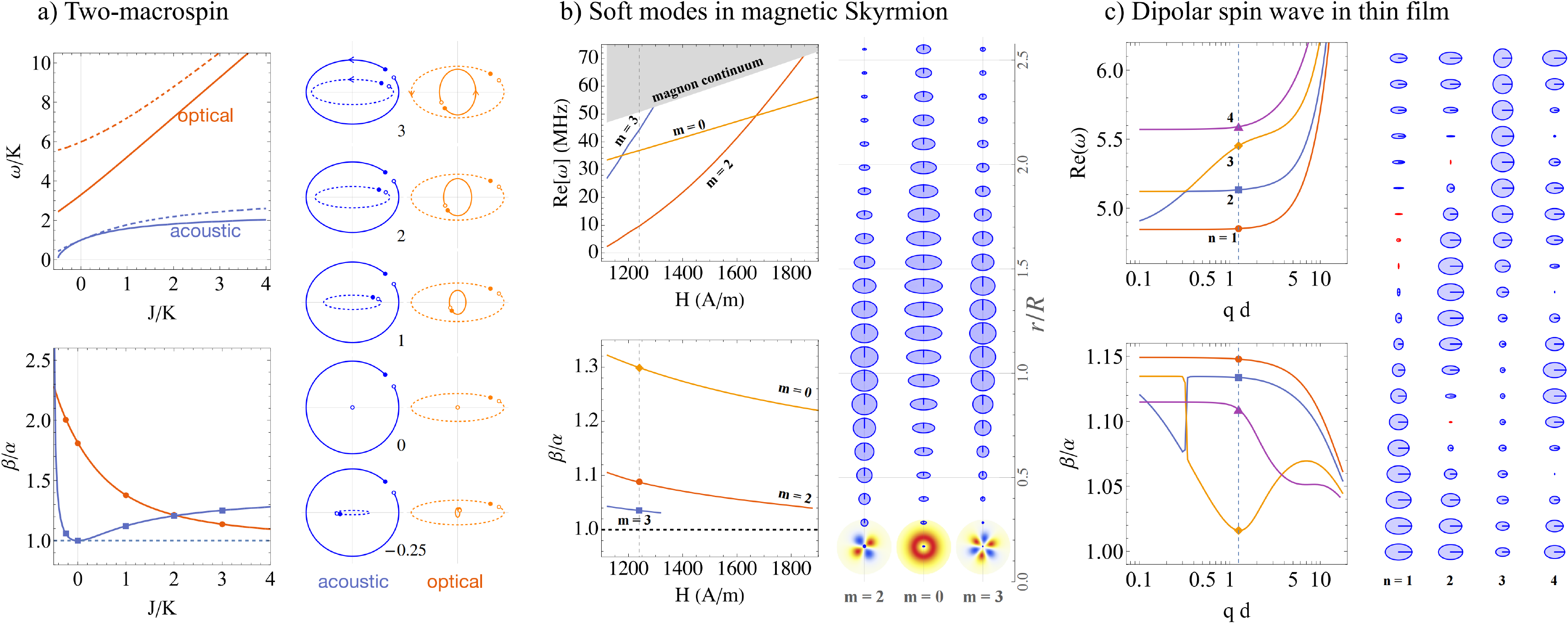}
    \caption{The simulated eigenfrequencies, dissipation rates, and spin wave profiles.
    (a) The acoustic and optical modes in the two-macrospin model. 
    Top panel: The real (solid) and imaginary (dashed, enlarged by $1/\alpha$) part of the eigenfrequencies for as function of exchange coupling $J$. 
    Bottom panel: dissipation rates obtained from the dispersions via \Eq{eqn:diss} (curves) and inferred from the spin wave profiles via \Eq{eqn:dpc} (points).
    Profile panel: The trajectories on $m_x$-$m_y$ projection for $\mb_1$ (solid) and $\mb_2$ (dashed) for $J/K = -0.25, 0, 1, 2, 3$. 
    (b) Same as (a) but for the soft modes in Skyrmion. The profiles (at $H = \SI{1240}{A/m}$) are along the radial direction. 
    (c) Same as (a) but for the dipolar spin wave in tangentially magnetized magnetic thin film with $\bq \perp \mb_0$ and thickness $d = \SI{305}{nm}$. The profiles are along the film thickness direction (at $qd = 1.25$). Blue and red orbits correspond to the right-handed and left-handed polarizations. The size of the ellipses is proportional to the square root of the real spin wave amplitudes.
    In all cases, $\alpha = 0.01$.
    }
    \label{fig:sim}
\end{figure*}

We now verify the polarization-dissipation connection \Eq{eqn:dpc} through several concrete examples, including coupled macrospins, magnetic domain walls, magnetic Skyrmions, and dipolar spin waves. Furthermore, we shall also demonstrate that this connection is not limited to ferromagnetic spin waves, but also applicable to antiferromagnetic spin waves. All simulation results in this paper are carried out using the micromagnetic module that we developed based on COMSOL Multiphysics, which has been applied to simulate ferromagnetic and antiferromagnetic spin waves in both time \cite{lan_spin-wave_2015,lan_antiferromagnetic_2017,yu_polarization-selective_2018,yu_magnetic_2020,yu_hopfield_2021} and frequency domain \cite{zhang_frequency-domain_2023}.

\emph{Coupled Macrospins - }
We consider a simple system consisting of two macrospins with free energy 
\begin{equation}
    \label{eqn:Fm1m2}
    F = - \frac{K}{2} \sum_{i=1,2}(\mb_i\cdot\hbz)^2 
    + \frac{K'}{2} (\mb_2\cdot\hby)^2 - J \mb_1 \cdot \mb_2,
\end{equation}
where $K$ is the uniaxial easy axis anisotropy along $\hbz$, $K'$ is the hard-axis anisotropy along $\hby$, and $J$ is the Heisenberg exchange coupling. The equilibrium magnetization of both sites point in $\hbz$: $\mb_1^0 = \mb_2^0 = \hbz$. Both macrospins have the same Gilbert damping parameter $\alpha$.
\Figure{fig:sim}(a) shows the real and imaginary eigenfrequencies of the two-macrospin system as function of the coupling strength $J$, along with magnetization trajectories for the two macrospins.
At $J = 0$, the mode 1, localized on $\mb_1$, has perfect circular polarization, thus its dissipation rate equals to $\alpha$. While, the mode 2, localized on $\mb_2$, has an elliptical polarization 
because of the additional hard-axis anisotropy, thus its dissipation rate is enhanced as given by \Eq{eqn:betar}. 
As $J$ becomes non-zero and positive (ferromagnetic coupling), the eigenmodes are no longer localized but encompass both macrospins. In the mean time, mode 1 (2) becomes more (less) elliptical. The dissipation rates  calculated from the complex eigenfrequencies (curves in the lower panel of \Figure{fig:sim}(a)) are in perfect agreement with the rates inferred from the polarization-dissipation connection \Eq{eqn:dpc} (points in lower panel of \Figure{fig:sim}(a)) based on the spin wave profiles given in right panel of \Figure{fig:sim}(a). 


\emph{Bipartite Antiferromagnet - }
Not only does the dissipation-polarization connection apply to ferromagnetic systems, but it also holds for antiferromagnetic systems. 
By letting let $J > 0, K' = 0$ in \Eq{eqn:Fm1m2}, we extend the coupled macrospin model above to the case of antiferromagnetic configuration with $\mb_1^0 = - \mb_2^0 = \hbz$.
The complex frequencies for the antiferromagnetic eigenmodes are \cite{keffer_theory_1952,wang_mechanism_2024}
$\omega_\pm = \pm \sqrt{K(K+2J)} + i\alpha (K+J)$.
And the dissipation rates are \cite{wang_mechanism_2024}
\begin{equation}
    \label{eqn:betaAF} 
    \beta = \alpha \frac{K+J}{\sqrt{K(K+2J)}}, 
\end{equation}
which is larger than the Gilbert damping constant $\alpha$. This enhancement can be understood using the dissipation-polarization connection in \Eq{eqn:dpc}. An important observation for antiferromagnetic spin wave is that
the magnetization from the two sublattices undergoes a circular precession with opposite handedness with respect to its local equilibrium magnetization directions: $\eta_1 = -\eta_2 = \pm 1$ for $\omega_\pm$ mode, and the precession cone angles $\theta_{1,2}$ have a ratio $(\theta_1/\theta_2)_\pm = [K+J \pm \sqrt{K(K+2J)}]/J$ \cite{keffer_theory_1952}.
Take the $\omega_+$ mode as an example, the opposite precession handedness means $\cosh(2r_1) = +1$ and $\cosh(2r_2) = -1$, and the directed area for the circular orbits of $\mb_1$ and $\mb_2$ are: $S_1 = +\pi\theta_1^2$ and $S_2 = -\pi\theta_2^2$, respectively. Therefore, according to \Eq{eqn:dpc}, the weighted polarization $\expval{\cosh(2r)} $ is 
\begin{equation*}
    \frac{(+1)\times (+\pi\theta_1^2) + (-1)\times(-\pi\theta_2^2)}{(+\pi\theta_1^2) + (-\pi\theta_2^2)}
    = \frac{K+J}{\sqrt{K(K+2J)}}, 
\end{equation*}
identical to the enhancement factor in \Eq{eqn:betaAF}, confirming the dissipation-polarization connection in antiferromagnet.

\emph{Domain wall - }
The dissipation-polarization connection also applies to spin wave excitations in complex magnetic textures.
An interesting case is the spin wave excitation in a magnetic domain wall with the magnetization rotating from one direction to another. 
One might anticipate that the inherently non-collinear structure of the domain wall would lead to an elliptical polarization of the spin waves, thus the dissipation rate would surpass the Gilbert damping constant. Contrary to this expectation, numerical simulations show that the linewidth of spin wave excitation in a magnetic domain wall is identical to the Gilbert damping constant. This intriguing finding indicates that, according to the dissipation-polarization connection \Eq{eqn:dpc}, the polarization of the spin wave remains perfectly circular as it traverses the domain wall.

\emph{Ferromagnetic Skyrmion - }
We now consider a ferromagnetic Skyrmion in the magnetic thin film with free energy 
\begin{align}
    F=\int \dd[2]{\br} &\left[-\frac{K}{2}(\mb \cdot \hbz)^2 +\frac{A}{2} (\nabla \mb)^2 \right. \nn
    &\left.-\frac{D}{2}\mb\cdot(\nabla \times \mb) - B \mb\cdot\hbz \right].
\end{align}
Here, we focus on the soft modes excitation in Skyrmion, including breathing, translational, and rotational modes. We do not anticipate these soft modes to possess circular polarization due to the inherent complexity of this non-collinear structure. \Figure{fig:sim}(b) shows the simulated results for the eigenfrequencies, the dissipation rates, and the corresponding spin wave profiles as function of external field $B$ applied perpendicular to the film.
The dissipation rates $\beta/\alpha$ (obtained by \Eq{eqn:diss}) shown as curves in the lower panel of \Figure{fig:sim}(b) are all greater than the Gilbert damping constant $\alpha$, implying the non-circular polarization for these modes (the right panel of \Figure{fig:sim}(b)). 
The dissipation rates inferred from the spin wave profiles are shown as the points in the lower panel of \Figure{fig:sim}(b). The exact agreement confirms the applicability of the dissipation-polarization connection \Eq{eqn:dpc} on the spin wave excitations with spatially varying amplitudes and ellipticities. 

\emph{Dipolar-exchange spin wave - }
The dissipation-polarization connection is not limited to local interactions such as the exchange interactions, but also applies to the non-local long range interactions such as the dipole-dipole interaction.
The dipolar interaction is intrinsically non-isotropic, leading to non-circular polarization.
We now consider the well studied dipolar-exchange spin wave in ferromagnetic thin films \cite{dewamesDipoleExchangeSpinWaves1970}, but we re-exam it from the viewpoint of its polarization and dissipation rate. 
We focus on the Damon-Eshbach mode with the equilibrium magnetization lying in the film plane and wave vector perpendicular to the magnetization. 
The simulated dispersion, dissipation rates (via \Eq{eqn:diss}), and the spin wave profiles are shown in \Figure{fig:sim}(c).
Since all profiles are elliptically polarized, according to the dissipation-polarization connection \Eq{eqn:dpc}, we immediately conclude that their dissipation rates shall all exceed $\alpha$, as seen in the lower panel of \Figure{fig:sim}(c). The points in the lower panel of \Figure{fig:sim}(c) are the dissipation rates inferred from the dissipation-polarization connection \Eq{eqn:dpc} based on the spin wave profiles. They match perfectly with the dissipation rates obtained from the dispersions via \Eq{eqn:diss}.
What's especially interesting for the dipolar-exchange spin wave is that the local precession can be even left-handed in a homogeneous ferromagnet \cite{dieterleCoherentExcitationHeterosymmetric2019}: the red profiles for modes $n = 1, 2$ indicating the left-handed motion. Because of this opposite polarization, the integral variable in the denominator of \Eq{eqn:dpc} is partly negative in these left-handed precession locations. The accounting of the signed weights is crucial for the agreement between dissipation rates obtained from dispersions \Eq{eqn:diss} and via the dissipation-polarization connection \Eq{eqn:dpc} for such situations.

\begin{figure}[t]
    \includegraphics[width=\columnwidth]{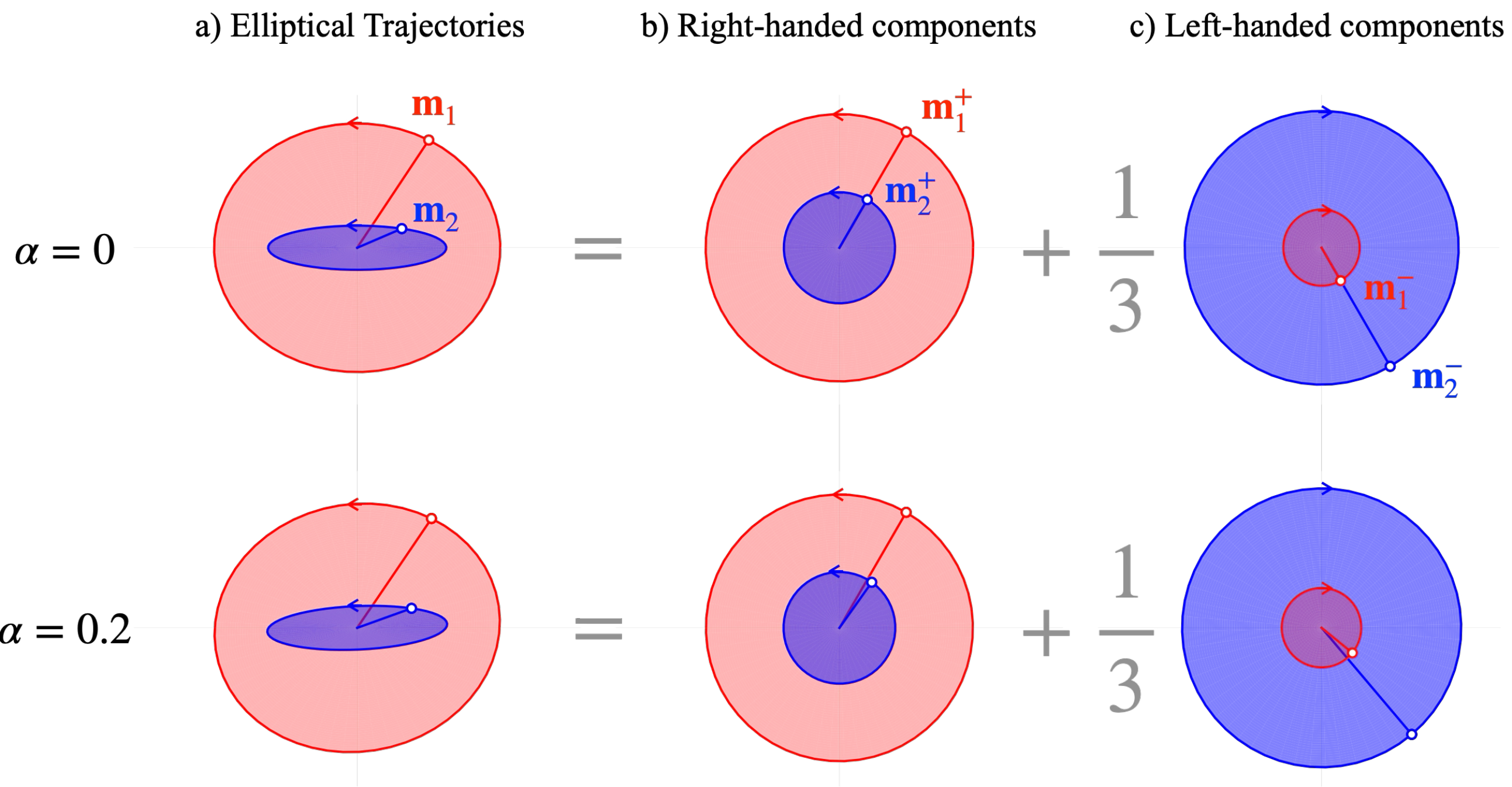}
    \caption{The magnetization trajectory for $\psi(t)$ in complex plane for the two-macrospin model. Left: The overall trajectory $\psi_{1,2}(t)$ for $\mb_1$ (red) and $\mb_2$ (blue). Middle: The right-handed component of the trajectory. Right: The left-handed component (enlarged by three times). The top row with $\alpha = 0$ shows no phase delay between two macrospins, and the bottom row with $\alpha = 0.2$ has a phase delay.}
    \label{fig:phase}
\end{figure}

\emph{Discussion - }
In the examples shown above, we observe that the spin wave polarization varies spatially, indicating that the dissipation rates also change across space in accordance with the dissipation-polarization connection \Eq{eqn:betar}. 
However, for an eigenmode to maintain its relative profile over time, the dissipation must occur uniformly.
This contradiction is resolved by a net energy flow from regions with lower dissipation rate to regions with higher rate, that is the fast-damping region helps dissipate part of the energy in the slow-damping region. In the (collinear) two-macrospin example above, such dissipation energy transfer is calculated by the energy transfer due to the exchange interaction over one period (see Appendix B): 
\begin{equation}
    J\int_0^T \dd{t} \dot{\mb}_1 \cdot \mb_2
    \propto \Im{m_1^+ {m_2^+}^* - m_1^- {m_2^-}^*}.
\end{equation}
This average energy transfer should vanish when Gilbert damping is turned off, indicating that the phase difference between the complex amplitudes at different locations must be either zero or $\pi$: $\arg{m_1^\pm} - \arg{m_2^\pm} = 0, \pi$ (see top row in \Figure{fig:phase} for the case of 0).
However, when Gilbert damping is turned on, the trajectories are modulated so that these amplitudes develops a small but critical phase difference $\arg{m_1^\pm} - \arg{m_2^\pm} \propto \alpha$ (see bottom row in \Figure{fig:phase}). Consequently, a net energy transfer between the two sites emerges, thus balancing their different dissipation rates.
A similar mechanism has been used to explain the enhanced dissipation rate in antiferromagnets by the present authors \cite{wang_mechanism_2024}.

Although the dissipation rate is influenced by the polarization of spin waves, it is important to note that this does not alter the Fluctuation-Dissipation Theorem (FDT) \cite{onsager_reciprocal_1931,kubo_fluctuation-dissipation_1966, brown_thermal_1963}, \ie the dissipation parameter in the FDT is still the original Gilbert damping parameter $\alpha$, rather than the enhanced rate $\beta$. 
We should also note that in this work we only consider the simplest local dissipation. Whether the dissipation-polarization connection applies to non-local dissipation such as $\mb_i \times \dot{\mb}_j$ with $i\neq j$ \cite{zhangGeneralizationLandauLifshitzGilbertEquation2009,tserkovnyakTransverseSpinDiffusion2009,yuanProperDissipativeTorques2019} requires further analysis. 

In conclusion, we have established a universal connection between the dissipation rate and the polarization of spin waves. This connection is exact for linear spin waves, and it holds for a wide range of systems, including ferromagnetic and antiferromagnetic spin waves, as well as spin waves in complex magnetic textures.

\bigskip
\emph{Acknowledgements. } 
This work was supported by 
National Natural Science Foundation of China (Grants No. 12474110),
the National Key Research and Development Program of China (Grant No. 2022YFA1403300),
the Innovation Program for Quantum Science and Technology (Grant No.2024ZD0300103),
and Shanghai Municipal Science and Technology Major Project (Grant No.2019SHZDZX01).

\bibliography{ref}

\begin{thebibliography}{42}%
\makeatletter
\providecommand \@ifxundefined [1]{%
 \@ifx{#1\undefined}
}%
\providecommand \@ifnum [1]{%
 \ifnum #1\expandafter \@firstoftwo
 \else \expandafter \@secondoftwo
 \fi
}%
\providecommand \@ifx [1]{%
 \ifx #1\expandafter \@firstoftwo
 \else \expandafter \@secondoftwo
 \fi
}%
\providecommand \natexlab [1]{#1}%
\providecommand \enquote  [1]{``#1''}%
\providecommand \bibnamefont  [1]{#1}%
\providecommand \bibfnamefont [1]{#1}%
\providecommand \citenamefont [1]{#1}%
\providecommand \href@noop [0]{\@secondoftwo}%
\providecommand \href [0]{\begingroup \@sanitize@url \@href}%
\providecommand \@href[1]{\@@startlink{#1}\@@href}%
\providecommand \@@href[1]{\endgroup#1\@@endlink}%
\providecommand \@sanitize@url [0]{\catcode `\\12\catcode `\$12\catcode
  `\&12\catcode `\#12\catcode `\^12\catcode `\_12\catcode `\%12\relax}%
\providecommand \@@startlink[1]{}%
\providecommand \@@endlink[0]{}%
\providecommand \url  [0]{\begingroup\@sanitize@url \@url }%
\providecommand \@url [1]{\endgroup\@href {#1}{\urlprefix }}%
\providecommand \urlprefix  [0]{URL }%
\providecommand \Eprint [0]{\href }%
\providecommand \doibase [0]{https://doi.org/}%
\providecommand \selectlanguage [0]{\@gobble}%
\providecommand \bibinfo  [0]{\@secondoftwo}%
\providecommand \bibfield  [0]{\@secondoftwo}%
\providecommand \translation [1]{[#1]}%
\providecommand \BibitemOpen [0]{}%
\providecommand \bibitemStop [0]{}%
\providecommand \bibitemNoStop [0]{.\EOS\space}%
\providecommand \EOS [0]{\spacefactor3000\relax}%
\providecommand \BibitemShut  [1]{\csname bibitem#1\endcsname}%
\let\auto@bib@innerbib\@empty
\bibitem [{\citenamefont {Stancil}\ and\ \citenamefont
  {Prabhakar}(2009)}]{stancil_spin_2009}%
  \BibitemOpen
  \bibfield  {author} {\bibinfo {author} {\bibfnamefont {D.~D.}\ \bibnamefont
  {Stancil}}\ and\ \bibinfo {author} {\bibfnamefont {A.}~\bibnamefont
  {Prabhakar}},\ }\href@noop {} {\emph {\bibinfo {title} {Spin {Waves}}}}\
  (\bibinfo  {publisher} {Springer},\ \bibinfo {address} {Office},\ \bibinfo
  {year} {2009})\BibitemShut {NoStop}%
\bibitem [{\citenamefont {Chumak}\ \emph {et~al.}(2021)\citenamefont {Chumak},
  \citenamefont {Kabos}, \citenamefont {Wu}, \citenamefont {Abert},
  \citenamefont {Adelmann}, \citenamefont {Adeyeye}, \citenamefont {{\r
  A}kerman}, \citenamefont {Aliev}, \citenamefont {Anane}, \citenamefont
  {Awad}, \citenamefont {Back}, \citenamefont {Barman}, \citenamefont {Bauer},
  \citenamefont {Becherer}, \citenamefont {Beginin}, \citenamefont
  {Bittencourt}, \citenamefont {Blanter}, \citenamefont {Bortolotti},
  \citenamefont {Boventer}, \citenamefont {Bozhko}, \citenamefont {Bunyaev},
  \citenamefont {Carmiggelt}, \citenamefont {Cheenikundil}, \citenamefont
  {Ciubotaru}, \citenamefont {Cotofana}, \citenamefont {Csaba}, \citenamefont
  {Dobrovolskiy}, \citenamefont {Dubs}, \citenamefont {Elyasi}, \citenamefont
  {Fripp}, \citenamefont {Fulara}, \citenamefont {Golovchanskiy}, \citenamefont
  {Gonzalez-Ballestero}, \citenamefont {Graczyk}, \citenamefont {Grundler},
  \citenamefont {Gruszecki}, \citenamefont {Gubbiotti}, \citenamefont
  {Guslienko}, \citenamefont {Haldar}, \citenamefont {Hamdioui}, \citenamefont
  {Hertel}, \citenamefont {Hillebrands}, \citenamefont {Hioki}, \citenamefont
  {Houshang}, \citenamefont {Hu}, \citenamefont {Huebl}, \citenamefont {Huth},
  \citenamefont {Iacocca}, \citenamefont {Jungfleisch}, \citenamefont
  {Kakazei}, \citenamefont {Khitun}, \citenamefont {Khymyn}, \citenamefont
  {Kikkawa}, \citenamefont {Kl{\"a}ui}, \citenamefont {Klein}, \citenamefont
  {Klos}, \citenamefont {Knauer}, \citenamefont {Koraltan}, \citenamefont
  {Kostylev}, \citenamefont {Krawczyk}, \citenamefont {Krivorotov},
  \citenamefont {Kruglyak}, \citenamefont {Lachance-Quirion}, \citenamefont
  {Ladak}, \citenamefont {Lebrun}, \citenamefont {Li}, \citenamefont {Lindner},
  \citenamefont {Mac{\^e}do}, \citenamefont {Melkov}, \citenamefont
  {Mieszczak}, \citenamefont {Nakamura}, \citenamefont {Nembach}, \citenamefont
  {Nikitin}, \citenamefont {Nikitov}, \citenamefont {Novosad}, \citenamefont
  {Otalora}, \citenamefont {Otani}, \citenamefont {Papp}, \citenamefont
  {Pigeau}, \citenamefont {Pirro}, \citenamefont {Porod}, \citenamefont
  {Porrati}, \citenamefont {Qin}, \citenamefont {Rana}, \citenamefont
  {Reimann}, \citenamefont {Riente}, \citenamefont {Romero-Isart},
  \citenamefont {Ross}, \citenamefont {Sadovnikov}, \citenamefont {Saitoh},
  \citenamefont {Schmidt}, \citenamefont {Schultheiss}, \citenamefont
  {Schultheiss}, \citenamefont {Serga}, \citenamefont {Sharma}, \citenamefont
  {Shaw}, \citenamefont {Suess}, \citenamefont {Surzhenko}, \citenamefont
  {Szulc}, \citenamefont {TANIGUCHI}, \citenamefont {Urb{\'a}nek},
  \citenamefont {Usami}, \citenamefont {Ustinov}, \citenamefont {Van Der~Sar},
  \citenamefont {Van~Dijken}, \citenamefont {Vasyuchka}, \citenamefont {Verba},
  \citenamefont {Kusminskiy}, \citenamefont {Wang}, \citenamefont {Weides},
  \citenamefont {Weiler}, \citenamefont {Wolski},\ and\ \citenamefont
  {Zhang}}]{chumak_roadmap_2021}%
  \BibitemOpen
  \bibfield  {author} {\bibinfo {author} {\bibfnamefont {A.~V.}\ \bibnamefont
  {Chumak}}, \bibinfo {author} {\bibfnamefont {P.}~\bibnamefont {Kabos}},
  \bibinfo {author} {\bibfnamefont {M.}~\bibnamefont {Wu}}, \bibinfo {author}
  {\bibfnamefont {C.}~\bibnamefont {Abert}}, \bibinfo {author} {\bibfnamefont
  {C.}~\bibnamefont {Adelmann}}, \bibinfo {author} {\bibfnamefont
  {A.}~\bibnamefont {Adeyeye}}, \bibinfo {author} {\bibfnamefont
  {J.}~\bibnamefont {{\r A}kerman}}, \bibinfo {author} {\bibfnamefont {F.~G.}\
  \bibnamefont {Aliev}}, \bibinfo {author} {\bibfnamefont {A.}~\bibnamefont
  {Anane}}, \bibinfo {author} {\bibfnamefont {A.}~\bibnamefont {Awad}},
  \bibinfo {author} {\bibfnamefont {C.~H.}\ \bibnamefont {Back}}, \bibinfo
  {author} {\bibfnamefont {A.}~\bibnamefont {Barman}}, \bibinfo {author}
  {\bibfnamefont {G.~E.~W.}\ \bibnamefont {Bauer}}, \bibinfo {author}
  {\bibfnamefont {M.}~\bibnamefont {Becherer}}, \bibinfo {author}
  {\bibfnamefont {E.~N.}\ \bibnamefont {Beginin}}, \bibinfo {author}
  {\bibfnamefont {V.~A. S.~V.}\ \bibnamefont {Bittencourt}}, \bibinfo {author}
  {\bibfnamefont {Y.~M.}\ \bibnamefont {Blanter}}, \bibinfo {author}
  {\bibfnamefont {P.}~\bibnamefont {Bortolotti}}, \bibinfo {author}
  {\bibfnamefont {I.}~\bibnamefont {Boventer}}, \bibinfo {author}
  {\bibfnamefont {D.~A.}\ \bibnamefont {Bozhko}}, \bibinfo {author}
  {\bibfnamefont {S.}~\bibnamefont {Bunyaev}}, \bibinfo {author} {\bibfnamefont
  {J.~J.}\ \bibnamefont {Carmiggelt}}, \bibinfo {author} {\bibfnamefont
  {R.~R.}\ \bibnamefont {Cheenikundil}}, \bibinfo {author} {\bibfnamefont
  {F.}~\bibnamefont {Ciubotaru}}, \bibinfo {author} {\bibfnamefont
  {S.}~\bibnamefont {Cotofana}}, \bibinfo {author} {\bibfnamefont
  {G.}~\bibnamefont {Csaba}}, \bibinfo {author} {\bibfnamefont
  {O.}~\bibnamefont {Dobrovolskiy}}, \bibinfo {author} {\bibfnamefont
  {C.}~\bibnamefont {Dubs}}, \bibinfo {author} {\bibfnamefont {M.}~\bibnamefont
  {Elyasi}}, \bibinfo {author} {\bibfnamefont {K.~G.}\ \bibnamefont {Fripp}},
  \bibinfo {author} {\bibfnamefont {H.}~\bibnamefont {Fulara}}, \bibinfo
  {author} {\bibfnamefont {I.~A.}\ \bibnamefont {Golovchanskiy}}, \bibinfo
  {author} {\bibfnamefont {C.}~\bibnamefont {Gonzalez-Ballestero}}, \bibinfo
  {author} {\bibfnamefont {P.}~\bibnamefont {Graczyk}}, \bibinfo {author}
  {\bibfnamefont {D.}~\bibnamefont {Grundler}}, \bibinfo {author}
  {\bibfnamefont {P.}~\bibnamefont {Gruszecki}}, \bibinfo {author}
  {\bibfnamefont {G.}~\bibnamefont {Gubbiotti}}, \bibinfo {author}
  {\bibfnamefont {K.}~\bibnamefont {Guslienko}}, \bibinfo {author}
  {\bibfnamefont {A.~O.}\ \bibnamefont {Haldar}}, \bibinfo {author}
  {\bibfnamefont {S.}~\bibnamefont {Hamdioui}}, \bibinfo {author}
  {\bibfnamefont {R.}~\bibnamefont {Hertel}}, \bibinfo {author} {\bibfnamefont
  {B.}~\bibnamefont {Hillebrands}}, \bibinfo {author} {\bibfnamefont
  {T.}~\bibnamefont {Hioki}}, \bibinfo {author} {\bibfnamefont
  {A.}~\bibnamefont {Houshang}}, \bibinfo {author} {\bibfnamefont {C.-M.}\
  \bibnamefont {Hu}}, \bibinfo {author} {\bibfnamefont {H.}~\bibnamefont
  {Huebl}}, \bibinfo {author} {\bibfnamefont {M.}~\bibnamefont {Huth}},
  \bibinfo {author} {\bibfnamefont {E.~N.}\ \bibnamefont {Iacocca}}, \bibinfo
  {author} {\bibfnamefont {M.~B.}\ \bibnamefont {Jungfleisch}}, \bibinfo
  {author} {\bibfnamefont {G.}~\bibnamefont {Kakazei}}, \bibinfo {author}
  {\bibfnamefont {A.}~\bibnamefont {Khitun}}, \bibinfo {author} {\bibfnamefont
  {R.}~\bibnamefont {Khymyn}}, \bibinfo {author} {\bibfnamefont
  {T.}~\bibnamefont {Kikkawa}}, \bibinfo {author} {\bibfnamefont
  {M.}~\bibnamefont {Kl{\"a}ui}}, \bibinfo {author} {\bibfnamefont
  {O.}~\bibnamefont {Klein}}, \bibinfo {author} {\bibfnamefont {J.~W.}\
  \bibnamefont {Klos}}, \bibinfo {author} {\bibfnamefont {S.}~\bibnamefont
  {Knauer}}, \bibinfo {author} {\bibfnamefont {S.}~\bibnamefont {Koraltan}},
  \bibinfo {author} {\bibfnamefont {M.}~\bibnamefont {Kostylev}}, \bibinfo
  {author} {\bibfnamefont {M.}~\bibnamefont {Krawczyk}}, \bibinfo {author}
  {\bibfnamefont {I.~N.}\ \bibnamefont {Krivorotov}}, \bibinfo {author}
  {\bibfnamefont {V.~V.}\ \bibnamefont {Kruglyak}}, \bibinfo {author}
  {\bibfnamefont {D.}~\bibnamefont {Lachance-Quirion}}, \bibinfo {author}
  {\bibfnamefont {S.}~\bibnamefont {Ladak}}, \bibinfo {author} {\bibfnamefont
  {R.}~\bibnamefont {Lebrun}}, \bibinfo {author} {\bibfnamefont
  {Y.}~\bibnamefont {Li}}, \bibinfo {author} {\bibfnamefont {M.}~\bibnamefont
  {Lindner}}, \bibinfo {author} {\bibfnamefont {R.}~\bibnamefont {Mac{\^e}do}},
  \bibinfo {author} {\bibfnamefont {G.~A.}\ \bibnamefont {Melkov}}, \bibinfo
  {author} {\bibfnamefont {S.}~\bibnamefont {Mieszczak}}, \bibinfo {author}
  {\bibfnamefont {Y.}~\bibnamefont {Nakamura}}, \bibinfo {author}
  {\bibfnamefont {H.}~\bibnamefont {Nembach}}, \bibinfo {author} {\bibfnamefont
  {A.~A.}\ \bibnamefont {Nikitin}}, \bibinfo {author} {\bibfnamefont {S.~A.}\
  \bibnamefont {Nikitov}}, \bibinfo {author} {\bibfnamefont {V.}~\bibnamefont
  {Novosad}}, \bibinfo {author} {\bibfnamefont {J.}~\bibnamefont {Otalora}},
  \bibinfo {author} {\bibfnamefont {Y.}~\bibnamefont {Otani}}, \bibinfo
  {author} {\bibfnamefont {A.}~\bibnamefont {Papp}}, \bibinfo {author}
  {\bibfnamefont {B.}~\bibnamefont {Pigeau}}, \bibinfo {author} {\bibfnamefont
  {P.}~\bibnamefont {Pirro}}, \bibinfo {author} {\bibfnamefont
  {W.}~\bibnamefont {Porod}}, \bibinfo {author} {\bibfnamefont
  {F.}~\bibnamefont {Porrati}}, \bibinfo {author} {\bibfnamefont
  {H.}~\bibnamefont {Qin}}, \bibinfo {author} {\bibfnamefont {B.}~\bibnamefont
  {Rana}}, \bibinfo {author} {\bibfnamefont {T.}~\bibnamefont {Reimann}},
  \bibinfo {author} {\bibfnamefont {F.}~\bibnamefont {Riente}}, \bibinfo
  {author} {\bibfnamefont {O.}~\bibnamefont {Romero-Isart}}, \bibinfo {author}
  {\bibfnamefont {A.}~\bibnamefont {Ross}}, \bibinfo {author} {\bibfnamefont
  {A.~V.}\ \bibnamefont {Sadovnikov}}, \bibinfo {author} {\bibfnamefont
  {E.}~\bibnamefont {Saitoh}}, \bibinfo {author} {\bibfnamefont
  {G.}~\bibnamefont {Schmidt}}, \bibinfo {author} {\bibfnamefont
  {H.}~\bibnamefont {Schultheiss}}, \bibinfo {author} {\bibfnamefont
  {K.}~\bibnamefont {Schultheiss}}, \bibinfo {author} {\bibfnamefont {A.~A.}\
  \bibnamefont {Serga}}, \bibinfo {author} {\bibfnamefont {S.}~\bibnamefont
  {Sharma}}, \bibinfo {author} {\bibfnamefont {J.}~\bibnamefont {Shaw}},
  \bibinfo {author} {\bibfnamefont {D.}~\bibnamefont {Suess}}, \bibinfo
  {author} {\bibfnamefont {O.}~\bibnamefont {Surzhenko}}, \bibinfo {author}
  {\bibfnamefont {K.}~\bibnamefont {Szulc}}, \bibinfo {author} {\bibfnamefont
  {T.}~\bibnamefont {TANIGUCHI}}, \bibinfo {author} {\bibfnamefont
  {M.}~\bibnamefont {Urb{\'a}nek}}, \bibinfo {author} {\bibfnamefont
  {K.}~\bibnamefont {Usami}}, \bibinfo {author} {\bibfnamefont {A.~B.}\
  \bibnamefont {Ustinov}}, \bibinfo {author} {\bibfnamefont {T.}~\bibnamefont
  {Van Der~Sar}}, \bibinfo {author} {\bibfnamefont {S.}~\bibnamefont
  {Van~Dijken}}, \bibinfo {author} {\bibfnamefont {V.~I.}\ \bibnamefont
  {Vasyuchka}}, \bibinfo {author} {\bibfnamefont {R.}~\bibnamefont {Verba}},
  \bibinfo {author} {\bibfnamefont {S.~V.}\ \bibnamefont {Kusminskiy}},
  \bibinfo {author} {\bibfnamefont {Q.}~\bibnamefont {Wang}}, \bibinfo {author}
  {\bibfnamefont {M.}~\bibnamefont {Weides}}, \bibinfo {author} {\bibfnamefont
  {M.}~\bibnamefont {Weiler}}, \bibinfo {author} {\bibfnamefont {S.~P.}\
  \bibnamefont {Wolski}},\ and\ \bibinfo {author} {\bibfnamefont
  {X.}~\bibnamefont {Zhang}},\ }\bibfield  {title} {\bibinfo {title} {Roadmap
  on spin-wave computing concepts},\ }\href
  {https://hal.archives-ouvertes.fr/hal-03381543} {\bibfield  {journal}
  {\bibinfo  {journal} {IEEE Transactions on Quantum Engineering}\ } (\bibinfo
  {year} {2021})},\ \bibinfo {note} {publisher: IEEE}\BibitemShut {NoStop}%
\bibitem [{\citenamefont {Demokritov}\ and\ \citenamefont
  {Slavin}(2012)}]{demokritov_magnonics:_2012}%
  \BibitemOpen
  \bibfield  {author} {\bibinfo {author} {\bibfnamefont {S.~O.}\ \bibnamefont
  {Demokritov}}\ and\ \bibinfo {author} {\bibfnamefont {A.~N.}\ \bibnamefont
  {Slavin}},\ }\href@noop {} {\emph {\bibinfo {title} {Magnonics: {From}
  {Fundamentals} to {Applications}}}}\ (\bibinfo  {publisher} {Springer Science
  \& Business Media},\ \bibinfo {year} {2012})\ \bibinfo {note}
  {00000}\BibitemShut {NoStop}%
\bibitem [{\citenamefont {Yu}\ \emph {et~al.}(2021{\natexlab{a}})\citenamefont
  {Yu}, \citenamefont {Xiao},\ and\ \citenamefont
  {Schultheiss}}]{yu_magnetic_2021}%
  \BibitemOpen
  \bibfield  {author} {\bibinfo {author} {\bibfnamefont {H.}~\bibnamefont
  {Yu}}, \bibinfo {author} {\bibfnamefont {J.}~\bibnamefont {Xiao}},\ and\
  \bibinfo {author} {\bibfnamefont {H.}~\bibnamefont {Schultheiss}},\
  }\bibfield  {title} {\bibinfo {title} {Magnetic texture based magnonics},\
  }\href {https://doi.org/10.1016/j.physrep.2020.12.004} {\bibfield  {journal}
  {\bibinfo  {journal} {Physics Reports}\ }\textbf {\bibinfo {volume} {905}},\
  \bibinfo {pages} {1} (\bibinfo {year} {2021}{\natexlab{a}})}\BibitemShut
  {NoStop}%
\bibitem [{\citenamefont {Yuan}\ \emph {et~al.}(2022)\citenamefont {Yuan},
  \citenamefont {Cao}, \citenamefont {Kamra}, \citenamefont {Duine},\ and\
  \citenamefont {Yan}}]{yuan_quantum_2022}%
  \BibitemOpen
  \bibfield  {author} {\bibinfo {author} {\bibfnamefont {H.}~\bibnamefont
  {Yuan}}, \bibinfo {author} {\bibfnamefont {Y.}~\bibnamefont {Cao}}, \bibinfo
  {author} {\bibfnamefont {A.}~\bibnamefont {Kamra}}, \bibinfo {author}
  {\bibfnamefont {R.~A.}\ \bibnamefont {Duine}},\ and\ \bibinfo {author}
  {\bibfnamefont {P.}~\bibnamefont {Yan}},\ }\bibfield  {title} {\bibinfo
  {title} {Quantum magnonics: {When} magnon spintronics meets quantum
  information science},\ }\href {https://doi.org/10.1016/j.physrep.2022.03.002}
  {\bibfield  {journal} {\bibinfo  {journal} {Physics Reports}\ }\textbf
  {\bibinfo {volume} {965}},\ \bibinfo {pages} {1} (\bibinfo {year}
  {2022})}\BibitemShut {NoStop}%
\bibitem [{\citenamefont {Zare~Rameshti}\ \emph {et~al.}(2022)\citenamefont
  {Zare~Rameshti}, \citenamefont {Viola~Kusminskiy}, \citenamefont {Haigh},
  \citenamefont {Usami}, \citenamefont {Lachance-Quirion}, \citenamefont
  {Nakamura}, \citenamefont {Hu}, \citenamefont {Tang}, \citenamefont {Bauer},\
  and\ \citenamefont {Blanter}}]{zare_rameshti_cavity_2022}%
  \BibitemOpen
  \bibfield  {author} {\bibinfo {author} {\bibfnamefont {B.}~\bibnamefont
  {Zare~Rameshti}}, \bibinfo {author} {\bibfnamefont {S.}~\bibnamefont
  {Viola~Kusminskiy}}, \bibinfo {author} {\bibfnamefont {J.~A.}\ \bibnamefont
  {Haigh}}, \bibinfo {author} {\bibfnamefont {K.}~\bibnamefont {Usami}},
  \bibinfo {author} {\bibfnamefont {D.}~\bibnamefont {Lachance-Quirion}},
  \bibinfo {author} {\bibfnamefont {Y.}~\bibnamefont {Nakamura}}, \bibinfo
  {author} {\bibfnamefont {C.-M.}\ \bibnamefont {Hu}}, \bibinfo {author}
  {\bibfnamefont {H.~X.}\ \bibnamefont {Tang}}, \bibinfo {author}
  {\bibfnamefont {G.~E.~W.}\ \bibnamefont {Bauer}},\ and\ \bibinfo {author}
  {\bibfnamefont {Y.~M.}\ \bibnamefont {Blanter}},\ }\bibfield  {title}
  {\bibinfo {title} {Cavity magnonics},\ }\href
  {https://doi.org/10.1016/j.physrep.2022.06.001} {\bibfield  {journal}
  {\bibinfo  {journal} {Physics Reports}\ }\bibinfo {series} {Cavity
  {Magnonics}},\ \textbf {\bibinfo {volume} {979}},\ \bibinfo {pages} {1}
  (\bibinfo {year} {2022})}\BibitemShut {NoStop}%
\bibitem [{\citenamefont {Gilbert}(2004)}]{gilbert_phenomenological_2004}%
  \BibitemOpen
  \bibfield  {author} {\bibinfo {author} {\bibfnamefont {T.}~\bibnamefont
  {Gilbert}},\ }\bibfield  {title} {\bibinfo {title} {A phenomenological theory
  of damping in ferromagnetic materials},\ }\href@noop {} {\bibfield  {journal}
  {\bibinfo  {journal} {Magnetics, IEEE Transactions on}\ }\textbf {\bibinfo
  {volume} {40}},\ \bibinfo {pages} {3443} (\bibinfo {year}
  {2004})}\BibitemShut {NoStop}%
\bibitem [{\citenamefont {Maendl}\ \emph {et~al.}(2017)\citenamefont {Maendl},
  \citenamefont {Stasinopoulos},\ and\ \citenamefont
  {Grundler}}]{maendlSpinWavesLarge2017}%
  \BibitemOpen
  \bibfield  {author} {\bibinfo {author} {\bibfnamefont {S.}~\bibnamefont
  {Maendl}}, \bibinfo {author} {\bibfnamefont {I.}~\bibnamefont
  {Stasinopoulos}},\ and\ \bibinfo {author} {\bibfnamefont {D.}~\bibnamefont
  {Grundler}},\ }\bibfield  {title} {\bibinfo {title} {Spin waves with large
  decay length and few 100\,nm wavelengths in thin yttrium iron garnet grown at
  the wafer scale},\ }\href {https://doi.org/10.1063/1.4991520} {\bibfield
  {journal} {\bibinfo  {journal} {Applied Physics Letters}\ }\textbf {\bibinfo
  {volume} {111}},\ \bibinfo {pages} {012403} (\bibinfo {year}
  {2017})}\BibitemShut {NoStop}%
\bibitem [{\citenamefont {Qin}\ \emph {et~al.}(2018)\citenamefont {Qin},
  \citenamefont {H{\"a}m{\"a}l{\"a}inen}, \citenamefont {Arjas}, \citenamefont
  {Witteveen},\ and\ \citenamefont {{van
  Dijken}}}]{qinPropagatingSpinWaves2018}%
  \BibitemOpen
  \bibfield  {author} {\bibinfo {author} {\bibfnamefont {H.}~\bibnamefont
  {Qin}}, \bibinfo {author} {\bibfnamefont {S.~J.}\ \bibnamefont
  {H{\"a}m{\"a}l{\"a}inen}}, \bibinfo {author} {\bibfnamefont {K.}~\bibnamefont
  {Arjas}}, \bibinfo {author} {\bibfnamefont {J.}~\bibnamefont {Witteveen}},\
  and\ \bibinfo {author} {\bibfnamefont {S.}~\bibnamefont {{van Dijken}}},\
  }\bibfield  {title} {\bibinfo {title} {Propagating spin waves in
  nanometer-thick yttrium iron garnet films: {{Dependence}} on wave vector,
  magnetic field strength, and angle},\ }\href
  {https://doi.org/10.1103/PhysRevB.98.224422} {\bibfield  {journal} {\bibinfo
  {journal} {Physical Review B}\ }\textbf {\bibinfo {volume} {98}},\ \bibinfo
  {pages} {224422} (\bibinfo {year} {2018})}\BibitemShut {NoStop}%
\bibitem [{\citenamefont {Schmidt}\ \emph {et~al.}(2020)\citenamefont
  {Schmidt}, \citenamefont {Hauser}, \citenamefont {Trempler}, \citenamefont
  {Paleschke},\ and\ \citenamefont {Papaioannou}}]{schmidtUltraThinFilms2020}%
  \BibitemOpen
  \bibfield  {author} {\bibinfo {author} {\bibfnamefont {G.}~\bibnamefont
  {Schmidt}}, \bibinfo {author} {\bibfnamefont {C.}~\bibnamefont {Hauser}},
  \bibinfo {author} {\bibfnamefont {P.}~\bibnamefont {Trempler}}, \bibinfo
  {author} {\bibfnamefont {M.}~\bibnamefont {Paleschke}},\ and\ \bibinfo
  {author} {\bibfnamefont {E.~T.}\ \bibnamefont {Papaioannou}},\ }\bibfield
  {title} {\bibinfo {title} {Ultra {{Thin Films}} of {{Yttrium Iron Garnet}}
  with {{Very Low Damping}}: {{A Review}}},\ }\href
  {https://doi.org/10.1002/pssb.201900644} {\bibfield  {journal} {\bibinfo
  {journal} {physica status solidi (b)}\ }\textbf {\bibinfo {volume} {257}},\
  \bibinfo {pages} {1900644} (\bibinfo {year} {2020})}\BibitemShut {NoStop}%
\bibitem [{\citenamefont {Azzawi}\ \emph {et~al.}(2017)\citenamefont {Azzawi},
  \citenamefont {Hindmarch},\ and\ \citenamefont
  {Atkinson}}]{azzawiMagneticDampingPhenomena2017}%
  \BibitemOpen
  \bibfield  {author} {\bibinfo {author} {\bibfnamefont {S.}~\bibnamefont
  {Azzawi}}, \bibinfo {author} {\bibfnamefont {A.~T.}\ \bibnamefont
  {Hindmarch}},\ and\ \bibinfo {author} {\bibfnamefont {D.}~\bibnamefont
  {Atkinson}},\ }\bibfield  {title} {\bibinfo {title} {Magnetic damping
  phenomena in ferromagnetic thin-films and multilayers},\ }\href
  {https://doi.org/10.1088/1361-6463/aa8dad} {\bibfield  {journal} {\bibinfo
  {journal} {Journal of Physics D: Applied Physics}\ }\textbf {\bibinfo
  {volume} {50}},\ \bibinfo {pages} {473001} (\bibinfo {year}
  {2017})}\BibitemShut {NoStop}%
\bibitem [{\citenamefont {Gurevich}\ and\ \citenamefont
  {Melkov}(1996)}]{gurevichMagnetizationOscillationsWaves1996}%
  \BibitemOpen
  \bibfield  {author} {\bibinfo {author} {\bibfnamefont {A.~G.}\ \bibnamefont
  {Gurevich}}\ and\ \bibinfo {author} {\bibfnamefont {G.~A.}\ \bibnamefont
  {Melkov}},\ }\href@noop {} {\emph {\bibinfo {title} {Magnetization
  Oscillations and Waves}}}\ (\bibinfo  {publisher} {CRC Press},\ \bibinfo
  {address} {Boca Raton},\ \bibinfo {year} {1996})\BibitemShut {NoStop}%
\bibitem [{\citenamefont {Gomonay}\ and\ \citenamefont
  {Loktev}(2014)}]{gomonay_spintronics_2014}%
  \BibitemOpen
  \bibfield  {author} {\bibinfo {author} {\bibfnamefont {E.~V.}\ \bibnamefont
  {Gomonay}}\ and\ \bibinfo {author} {\bibfnamefont {V.~M.}\ \bibnamefont
  {Loktev}},\ }\bibfield  {title} {\bibinfo {title} {Spintronics of
  antiferromagnetic systems ({Review} {Article})},\ }\href
  {https://doi.org/10.1063/1.4862467} {\bibfield  {journal} {\bibinfo
  {journal} {Low Temperature Physics}\ }\textbf {\bibinfo {volume} {40}},\
  \bibinfo {pages} {17} (\bibinfo {year} {2014})}\BibitemShut {NoStop}%
\bibitem [{\citenamefont {Lan}\ \emph {et~al.}(2017{\natexlab{a}})\citenamefont
  {Lan}, \citenamefont {Yu},\ and\ \citenamefont
  {Xiao}}]{lanAntiferromagneticDomainWall2017}%
  \BibitemOpen
  \bibfield  {author} {\bibinfo {author} {\bibfnamefont {J.}~\bibnamefont
  {Lan}}, \bibinfo {author} {\bibfnamefont {W.}~\bibnamefont {Yu}},\ and\
  \bibinfo {author} {\bibfnamefont {J.}~\bibnamefont {Xiao}},\ }\bibfield
  {title} {\bibinfo {title} {Antiferromagnetic domain wall as spin wave
  polarizer and retarder},\ }\href {https://doi.org/10.1038/s41467-017-00265-5}
  {\bibfield  {journal} {\bibinfo  {journal} {Nature Communications}\ }\textbf
  {\bibinfo {volume} {8}},\ \bibinfo {pages} {178} (\bibinfo {year}
  {2017}{\natexlab{a}})}\BibitemShut {NoStop}%
\bibitem [{\citenamefont {Cheng}\ \emph {et~al.}(2016)\citenamefont {Cheng},
  \citenamefont {Daniels}, \citenamefont {Zhu},\ and\ \citenamefont
  {Xiao}}]{cheng_antiferromagnetic_2016}%
  \BibitemOpen
  \bibfield  {author} {\bibinfo {author} {\bibfnamefont {R.}~\bibnamefont
  {Cheng}}, \bibinfo {author} {\bibfnamefont {M.~W.}\ \bibnamefont {Daniels}},
  \bibinfo {author} {\bibfnamefont {J.-G.}\ \bibnamefont {Zhu}},\ and\ \bibinfo
  {author} {\bibfnamefont {D.}~\bibnamefont {Xiao}},\ }\bibfield  {title}
  {\bibinfo {title} {Antiferromagnetic {Spin} {Wave} {Field}-{Effect}
  {Transistor}},\ }\href {https://doi.org/10.1038/srep24223} {\bibfield
  {journal} {\bibinfo  {journal} {Scientific Reports}\ }\textbf {\bibinfo
  {volume} {6}},\ \bibinfo {pages} {24223} (\bibinfo {year}
  {2016})}\BibitemShut {NoStop}%
\bibitem [{\citenamefont {Lan}\ \emph {et~al.}(2017{\natexlab{b}})\citenamefont
  {Lan}, \citenamefont {Yu},\ and\ \citenamefont
  {Xiao}}]{lan_antiferromagnetic_2017}%
  \BibitemOpen
  \bibfield  {author} {\bibinfo {author} {\bibfnamefont {J.}~\bibnamefont
  {Lan}}, \bibinfo {author} {\bibfnamefont {W.}~\bibnamefont {Yu}},\ and\
  \bibinfo {author} {\bibfnamefont {J.}~\bibnamefont {Xiao}},\ }\bibfield
  {title} {\bibinfo {title} {Antiferromagnetic domain wall as spin wave
  polarizer and retarder},\ }\href {https://doi.org/10.1038/s41467-017-00265-5}
  {\bibfield  {journal} {\bibinfo  {journal} {Nature Communications}\ }\textbf
  {\bibinfo {volume} {8}},\ \bibinfo {pages} {178} (\bibinfo {year}
  {2017}{\natexlab{b}})}\BibitemShut {NoStop}%
\bibitem [{\citenamefont {Yu}\ \emph {et~al.}(2018{\natexlab{a}})\citenamefont
  {Yu}, \citenamefont {Lan},\ and\ \citenamefont
  {Xiao}}]{yuPolarizationselectiveSpinWave2018a}%
  \BibitemOpen
  \bibfield  {author} {\bibinfo {author} {\bibfnamefont {W.}~\bibnamefont
  {Yu}}, \bibinfo {author} {\bibfnamefont {J.}~\bibnamefont {Lan}},\ and\
  \bibinfo {author} {\bibfnamefont {J.}~\bibnamefont {Xiao}},\ }\bibfield
  {title} {\bibinfo {title} {Polarization-selective spin wave driven
  domain-wall motion in antiferromagnets},\ }\href
  {https://doi.org/10.1103/PhysRevB.98.144422} {\bibfield  {journal} {\bibinfo
  {journal} {Physical Review B}\ }\textbf {\bibinfo {volume} {98}},\ \bibinfo
  {pages} {144422} (\bibinfo {year} {2018}{\natexlab{a}})}\BibitemShut
  {NoStop}%
\bibitem [{\citenamefont {Yu}\ \emph {et~al.}(2020)\citenamefont {Yu},
  \citenamefont {Lan},\ and\ \citenamefont {Xiao}}]{yu_magnetic_2020}%
  \BibitemOpen
  \bibfield  {author} {\bibinfo {author} {\bibfnamefont {W.}~\bibnamefont
  {Yu}}, \bibinfo {author} {\bibfnamefont {J.}~\bibnamefont {Lan}},\ and\
  \bibinfo {author} {\bibfnamefont {J.}~\bibnamefont {Xiao}},\ }\bibfield
  {title} {\bibinfo {title} {Magnetic {Logic} {Gate} {Based} on {Polarized}
  {Spin} {Waves}},\ }\href {https://doi.org/10.1103/PhysRevApplied.13.024055}
  {\bibfield  {journal} {\bibinfo  {journal} {Physical Review Applied}\
  }\textbf {\bibinfo {volume} {13}},\ \bibinfo {pages} {024055} (\bibinfo
  {year} {2020})}\BibitemShut {NoStop}%
\bibitem [{\citenamefont {Han}\ \emph {et~al.}(2020)\citenamefont {Han},
  \citenamefont {Zhang}, \citenamefont {Bi}, \citenamefont {Fan}, \citenamefont
  {Safi}, \citenamefont {Xiang}, \citenamefont {Finley}, \citenamefont {Fu},
  \citenamefont {Cheng},\ and\ \citenamefont
  {Liu}}]{han_birefringence-like_2020}%
  \BibitemOpen
  \bibfield  {author} {\bibinfo {author} {\bibfnamefont {J.}~\bibnamefont
  {Han}}, \bibinfo {author} {\bibfnamefont {P.}~\bibnamefont {Zhang}}, \bibinfo
  {author} {\bibfnamefont {Z.}~\bibnamefont {Bi}}, \bibinfo {author}
  {\bibfnamefont {Y.}~\bibnamefont {Fan}}, \bibinfo {author} {\bibfnamefont
  {T.~S.}\ \bibnamefont {Safi}}, \bibinfo {author} {\bibfnamefont
  {J.}~\bibnamefont {Xiang}}, \bibinfo {author} {\bibfnamefont
  {J.}~\bibnamefont {Finley}}, \bibinfo {author} {\bibfnamefont
  {L.}~\bibnamefont {Fu}}, \bibinfo {author} {\bibfnamefont {R.}~\bibnamefont
  {Cheng}},\ and\ \bibinfo {author} {\bibfnamefont {L.}~\bibnamefont {Liu}},\
  }\bibfield  {title} {\bibinfo {title} {Birefringence-like spin transport via
  linearly polarized antiferromagnetic magnons},\ }\href
  {https://doi.org/10.1038/s41565-020-0703-8} {\bibfield  {journal} {\bibinfo
  {journal} {Nature Nanotechnology}\ ,\ \bibinfo {pages} {1}} (\bibinfo {year}
  {2020})},\ \bibinfo {note} {publisher: Nature Publishing Group}\BibitemShut
  {NoStop}%
\bibitem [{\citenamefont {Kostylev}\ \emph {et~al.}(2005)\citenamefont
  {Kostylev}, \citenamefont {Serga}, \citenamefont {Schneider}, \citenamefont
  {Leven},\ and\ \citenamefont
  {Hillebrands}}]{kostylevSpinwaveLogicalGates2005}%
  \BibitemOpen
  \bibfield  {author} {\bibinfo {author} {\bibfnamefont {M.~P.}\ \bibnamefont
  {Kostylev}}, \bibinfo {author} {\bibfnamefont {A.~A.}\ \bibnamefont {Serga}},
  \bibinfo {author} {\bibfnamefont {T.}~\bibnamefont {Schneider}}, \bibinfo
  {author} {\bibfnamefont {B.}~\bibnamefont {Leven}},\ and\ \bibinfo {author}
  {\bibfnamefont {B.}~\bibnamefont {Hillebrands}},\ }\bibfield  {title}
  {\bibinfo {title} {Spin-wave logical gates},\ }\href
  {https://doi.org/10.1063/1.2089147} {\bibfield  {journal} {\bibinfo
  {journal} {Applied Physics Letters}\ }\textbf {\bibinfo {volume} {87}},\
  \bibinfo {pages} {153501} (\bibinfo {year} {2005})}\BibitemShut {NoStop}%
\bibitem [{\citenamefont {Schneider}\ \emph {et~al.}(2008)\citenamefont
  {Schneider}, \citenamefont {Serga}, \citenamefont {Leven}, \citenamefont
  {Hillebrands}, \citenamefont {Stamps},\ and\ \citenamefont
  {Kostylev}}]{schneider_realization_2008}%
  \BibitemOpen
  \bibfield  {author} {\bibinfo {author} {\bibfnamefont {T.}~\bibnamefont
  {Schneider}}, \bibinfo {author} {\bibfnamefont {A.~A.}\ \bibnamefont
  {Serga}}, \bibinfo {author} {\bibfnamefont {B.}~\bibnamefont {Leven}},
  \bibinfo {author} {\bibfnamefont {B.}~\bibnamefont {Hillebrands}}, \bibinfo
  {author} {\bibfnamefont {R.~L.}\ \bibnamefont {Stamps}},\ and\ \bibinfo
  {author} {\bibfnamefont {M.~P.}\ \bibnamefont {Kostylev}},\ }\bibfield
  {title} {\bibinfo {title} {Realization of spin-wave logic gates},\ }\href
  {https://doi.org/10.1063/1.2834714} {\bibfield  {journal} {\bibinfo
  {journal} {Applied Physics Letters}\ }\textbf {\bibinfo {volume} {92}},\
  \bibinfo {pages} {022505} (\bibinfo {year} {2008})}\BibitemShut {NoStop}%
\bibitem [{\citenamefont {Kruglyak}\ \emph {et~al.}(2010)\citenamefont
  {Kruglyak}, \citenamefont {Demokritov},\ and\ \citenamefont
  {Grundler}}]{kruglyakMagnonics2010}%
  \BibitemOpen
  \bibfield  {author} {\bibinfo {author} {\bibfnamefont {V.~V.}\ \bibnamefont
  {Kruglyak}}, \bibinfo {author} {\bibfnamefont {S.~O.}\ \bibnamefont
  {Demokritov}},\ and\ \bibinfo {author} {\bibfnamefont {D.}~\bibnamefont
  {Grundler}},\ }\bibfield  {title} {\bibinfo {title} {Magnonics},\ }\href
  {https://doi.org/10.1088/0022-3727/43/26/264001} {\bibfield  {journal}
  {\bibinfo  {journal} {Journal of Physics D: Applied Physics}\ }\textbf
  {\bibinfo {volume} {43}},\ \bibinfo {pages} {264001} (\bibinfo {year}
  {2010})}\BibitemShut {NoStop}%
\bibitem [{\citenamefont {Khitun}\ and\ \citenamefont
  {Wang}(2011)}]{khitunNonvolatileMagnonicLogic2011}%
  \BibitemOpen
  \bibfield  {author} {\bibinfo {author} {\bibfnamefont {A.}~\bibnamefont
  {Khitun}}\ and\ \bibinfo {author} {\bibfnamefont {K.~L.}\ \bibnamefont
  {Wang}},\ }\bibfield  {title} {\bibinfo {title} {Non-volatile magnonic logic
  circuits engineering},\ }\href {https://doi.org/10.1063/1.3609062} {\bibfield
   {journal} {\bibinfo  {journal} {Journal of Applied Physics}\ }\textbf
  {\bibinfo {volume} {110}},\ \bibinfo {pages} {034306} (\bibinfo {year}
  {2011})}\BibitemShut {NoStop}%
\bibitem [{\citenamefont {Mahmoud}\ \emph {et~al.}(2020)\citenamefont
  {Mahmoud}, \citenamefont {Ciubotaru}, \citenamefont {Vanderveken},
  \citenamefont {Chumak}, \citenamefont {Hamdioui}, \citenamefont {Adelmann},\
  and\ \citenamefont {Cotofana}}]{mahmoud_introduction_2020}%
  \BibitemOpen
  \bibfield  {author} {\bibinfo {author} {\bibfnamefont {A.}~\bibnamefont
  {Mahmoud}}, \bibinfo {author} {\bibfnamefont {F.}~\bibnamefont {Ciubotaru}},
  \bibinfo {author} {\bibfnamefont {F.}~\bibnamefont {Vanderveken}}, \bibinfo
  {author} {\bibfnamefont {A.~V.}\ \bibnamefont {Chumak}}, \bibinfo {author}
  {\bibfnamefont {S.}~\bibnamefont {Hamdioui}}, \bibinfo {author}
  {\bibfnamefont {C.}~\bibnamefont {Adelmann}},\ and\ \bibinfo {author}
  {\bibfnamefont {S.}~\bibnamefont {Cotofana}},\ }\bibfield  {title} {\bibinfo
  {title} {Introduction to spin wave computing},\ }\href
  {https://doi.org/10.1063/5.0019328} {\bibfield  {journal} {\bibinfo
  {journal} {Journal of Applied Physics}\ }\textbf {\bibinfo {volume} {128}},\
  \bibinfo {pages} {161101} (\bibinfo {year} {2020})}\BibitemShut {NoStop}%
\bibitem [{\citenamefont {Kambersky}\ and\ \citenamefont
  {Patton}(1975)}]{kambersky_spin-wave_1975}%
  \BibitemOpen
  \bibfield  {author} {\bibinfo {author} {\bibfnamefont {V.}~\bibnamefont
  {Kambersky}}\ and\ \bibinfo {author} {\bibfnamefont {C.~E.}\ \bibnamefont
  {Patton}},\ }\bibfield  {title} {\bibinfo {title} {Spin-wave relaxation and
  phenomenological damping in ferromagnetic resonance},\ }\href
  {https://doi.org/10.1103/PhysRevB.11.2668} {\bibfield  {journal} {\bibinfo
  {journal} {Physical Review B}\ }\textbf {\bibinfo {volume} {11}},\ \bibinfo
  {pages} {2668} (\bibinfo {year} {1975})},\ \bibinfo {note} {publisher:
  American Physical Society}\BibitemShut {NoStop}%
\bibitem [{\citenamefont {Puszkarski}(1979)}]{puszkarski_theory_1979}%
  \BibitemOpen
  \bibfield  {author} {\bibinfo {author} {\bibfnamefont {H.}~\bibnamefont
  {Puszkarski}},\ }\bibfield  {title} {\bibinfo {title} {Theory of surface
  states in spin wave resonance},\ }\href
  {https://doi.org/10.1016/0079-6816(79)90013-3} {\bibfield  {journal}
  {\bibinfo  {journal} {Progress in Surface Science}\ }\textbf {\bibinfo
  {volume} {9}},\ \bibinfo {pages} {191} (\bibinfo {year} {1979})}\BibitemShut
  {NoStop}%
\bibitem [{\citenamefont {R{\'o}zsa}\ \emph {et~al.}(2018)\citenamefont
  {R{\'o}zsa}, \citenamefont {Hagemeister}, \citenamefont {Vedmedenko},\ and\
  \citenamefont {Wiesendanger}}]{rozsa_effective_2018}%
  \BibitemOpen
  \bibfield  {author} {\bibinfo {author} {\bibfnamefont {L.}~\bibnamefont
  {R{\'o}zsa}}, \bibinfo {author} {\bibfnamefont {J.}~\bibnamefont
  {Hagemeister}}, \bibinfo {author} {\bibfnamefont {E.~Y.}\ \bibnamefont
  {Vedmedenko}},\ and\ \bibinfo {author} {\bibfnamefont {R.}~\bibnamefont
  {Wiesendanger}},\ }\bibfield  {title} {\bibinfo {title} {Effective damping
  enhancement in noncollinear spin structures},\ }\href
  {https://doi.org/10.1103/PhysRevB.98.100404} {\bibfield  {journal} {\bibinfo
  {journal} {Physical Review B}\ }\textbf {\bibinfo {volume} {98}},\ \bibinfo
  {pages} {100404} (\bibinfo {year} {2018})},\ \bibinfo {note} {publisher:
  American Physical Society}\BibitemShut {NoStop}%
\bibitem [{\citenamefont {Patton}(1968)}]{patton_linewidth_1968}%
  \BibitemOpen
  \bibfield  {author} {\bibinfo {author} {\bibfnamefont {C.~E.}\ \bibnamefont
  {Patton}},\ }\bibfield  {title} {\bibinfo {title} {Linewidth and {Relaxation}
  {Processes} for the {Main} {Resonance} in the {Spin}-{Wave} {Spectra} of
  {Ni}{\textendash}{Fe} {Alloy} {Films}},\ }\href
  {https://doi.org/10.1063/1.1656733} {\bibfield  {journal} {\bibinfo
  {journal} {Journal of Applied Physics}\ }\textbf {\bibinfo {volume} {39}},\
  \bibinfo {pages} {3060} (\bibinfo {year} {1968})}\BibitemShut {NoStop}%
\bibitem [{\citenamefont {Lan}\ \emph {et~al.}(2015)\citenamefont {Lan},
  \citenamefont {Yu}, \citenamefont {Wu},\ and\ \citenamefont
  {Xiao}}]{lan_spin-wave_2015}%
  \BibitemOpen
  \bibfield  {author} {\bibinfo {author} {\bibfnamefont {J.}~\bibnamefont
  {Lan}}, \bibinfo {author} {\bibfnamefont {W.}~\bibnamefont {Yu}}, \bibinfo
  {author} {\bibfnamefont {R.}~\bibnamefont {Wu}},\ and\ \bibinfo {author}
  {\bibfnamefont {J.}~\bibnamefont {Xiao}},\ }\bibfield  {title} {\bibinfo
  {title} {Spin-{Wave} {Diode}},\ }\href
  {https://doi.org/10.1103/PhysRevX.5.041049} {\bibfield  {journal} {\bibinfo
  {journal} {Physical Review X}\ }\textbf {\bibinfo {volume} {5}},\ \bibinfo
  {pages} {041049} (\bibinfo {year} {2015})}\BibitemShut {NoStop}%
\bibitem [{\citenamefont {Yu}\ \emph {et~al.}(2018{\natexlab{b}})\citenamefont
  {Yu}, \citenamefont {Lan},\ and\ \citenamefont
  {Xiao}}]{yu_polarization-selective_2018}%
  \BibitemOpen
  \bibfield  {author} {\bibinfo {author} {\bibfnamefont {W.}~\bibnamefont
  {Yu}}, \bibinfo {author} {\bibfnamefont {J.}~\bibnamefont {Lan}},\ and\
  \bibinfo {author} {\bibfnamefont {J.}~\bibnamefont {Xiao}},\ }\bibfield
  {title} {\bibinfo {title} {Polarization-selective spin wave driven
  domain-wall motion in antiferromagnets},\ }\href
  {https://doi.org/10.1103/PhysRevB.98.144422} {\bibfield  {journal} {\bibinfo
  {journal} {Physical Review B}\ }\textbf {\bibinfo {volume} {98}},\ \bibinfo
  {pages} {144422} (\bibinfo {year} {2018}{\natexlab{b}})}\BibitemShut
  {NoStop}%
\bibitem [{\citenamefont {Yu}\ \emph {et~al.}(2021{\natexlab{b}})\citenamefont
  {Yu}, \citenamefont {Xiao},\ and\ \citenamefont {Bauer}}]{yu_hopfield_2021}%
  \BibitemOpen
  \bibfield  {author} {\bibinfo {author} {\bibfnamefont {W.}~\bibnamefont
  {Yu}}, \bibinfo {author} {\bibfnamefont {J.}~\bibnamefont {Xiao}},\ and\
  \bibinfo {author} {\bibfnamefont {G.~E.~W.}\ \bibnamefont {Bauer}},\
  }\bibfield  {title} {\bibinfo {title} {Hopfield neural network in magnetic
  textures with intrinsic {Hebbian} learning},\ }\href
  {https://doi.org/10.1103/PhysRevB.104.L180405} {\bibfield  {journal}
  {\bibinfo  {journal} {Physical Review B}\ }\textbf {\bibinfo {volume}
  {104}},\ \bibinfo {pages} {L180405} (\bibinfo {year} {2021}{\natexlab{b}})},\
  \bibinfo {note} {publisher: American Physical Society}\BibitemShut {NoStop}%
\bibitem [{\citenamefont {Zhang}\ \emph {et~al.}(2023)\citenamefont {Zhang},
  \citenamefont {Yu}, \citenamefont {Chen},\ and\ \citenamefont
  {Xiao}}]{zhang_frequency-domain_2023}%
  \BibitemOpen
  \bibfield  {author} {\bibinfo {author} {\bibfnamefont {J.}~\bibnamefont
  {Zhang}}, \bibinfo {author} {\bibfnamefont {W.}~\bibnamefont {Yu}}, \bibinfo
  {author} {\bibfnamefont {X.}~\bibnamefont {Chen}},\ and\ \bibinfo {author}
  {\bibfnamefont {J.}~\bibnamefont {Xiao}},\ }\bibfield  {title} {\bibinfo
  {title} {A frequency-domain micromagnetic simulation module based on {COMSOL}
  {Multiphysics}},\ }\href {https://doi.org/10.1063/5.0143262} {\bibfield
  {journal} {\bibinfo  {journal} {AIP Advances}\ }\textbf {\bibinfo {volume}
  {13}},\ \bibinfo {pages} {055108} (\bibinfo {year} {2023})}\BibitemShut
  {NoStop}%
\bibitem [{\citenamefont {Keffer}\ and\ \citenamefont
  {Kittel}(1952)}]{keffer_theory_1952}%
  \BibitemOpen
  \bibfield  {author} {\bibinfo {author} {\bibfnamefont {F.}~\bibnamefont
  {Keffer}}\ and\ \bibinfo {author} {\bibfnamefont {C.}~\bibnamefont
  {Kittel}},\ }\bibfield  {title} {\bibinfo {title} {Theory of
  {Antiferromagnetic} {Resonance}},\ }\href
  {https://doi.org/10.1103/PhysRev.85.329} {\bibfield  {journal} {\bibinfo
  {journal} {Physical Review}\ }\textbf {\bibinfo {volume} {85}},\ \bibinfo
  {pages} {329} (\bibinfo {year} {1952})},\ \bibinfo {note} {00207}\BibitemShut
  {NoStop}%
\bibitem [{\citenamefont {Wang}\ and\ \citenamefont
  {Xiao}(2024)}]{wang_mechanism_2024}%
  \BibitemOpen
  \bibfield  {author} {\bibinfo {author} {\bibfnamefont {Y.}~\bibnamefont
  {Wang}}\ and\ \bibinfo {author} {\bibfnamefont {J.}~\bibnamefont {Xiao}},\
  }\bibfield  {title} {\bibinfo {title} {Mechanism for broadened linewidth in
  antiferromagnetic resonance},\ }\href
  {https://doi.org/10.1103/PhysRevB.110.134409} {\bibfield  {journal} {\bibinfo
   {journal} {Physical Review B}\ }\textbf {\bibinfo {volume} {110}},\ \bibinfo
  {pages} {134409} (\bibinfo {year} {2024})},\ \bibinfo {note} {publisher:
  American Physical Society}\BibitemShut {NoStop}%
\bibitem [{\citenamefont {De~Wames}\ and\ \citenamefont
  {Wolfram}(1970)}]{dewamesDipoleExchangeSpinWaves1970}%
  \BibitemOpen
  \bibfield  {author} {\bibinfo {author} {\bibfnamefont {R.~E.}\ \bibnamefont
  {De~Wames}}\ and\ \bibinfo {author} {\bibfnamefont {T.}~\bibnamefont
  {Wolfram}},\ }\bibfield  {title} {\bibinfo {title} {Dipole-{{Exchange Spin
  Waves}} in {{Ferromagnetic Films}}},\ }\href
  {https://doi.org/10.1063/1.1659049} {\bibfield  {journal} {\bibinfo
  {journal} {Journal of Applied Physics}\ }\textbf {\bibinfo {volume} {41}},\
  \bibinfo {pages} {987} (\bibinfo {year} {1970})}\BibitemShut {NoStop}%
\bibitem [{\citenamefont {Dieterle}\ \emph {et~al.}(2019)\citenamefont
  {Dieterle}, \citenamefont {F{\"o}rster}, \citenamefont {Stoll}, \citenamefont
  {Semisalova}, \citenamefont {Finizio}, \citenamefont {Gangwar}, \citenamefont
  {Weigand}, \citenamefont {Noske}, \citenamefont {F{\"a}hnle}, \citenamefont
  {Bykova}, \citenamefont {Gr{\"a}fe}, \citenamefont {Bozhko}, \citenamefont
  {{Musiienko-Shmarova}}, \citenamefont {Tiberkevich}, \citenamefont {Slavin},
  \citenamefont {Back}, \citenamefont {Raabe}, \citenamefont {Sch{\"u}tz},\
  and\ \citenamefont {Wintz}}]{dieterleCoherentExcitationHeterosymmetric2019}%
  \BibitemOpen
  \bibfield  {author} {\bibinfo {author} {\bibfnamefont {G.}~\bibnamefont
  {Dieterle}}, \bibinfo {author} {\bibfnamefont {J.}~\bibnamefont
  {F{\"o}rster}}, \bibinfo {author} {\bibfnamefont {H.}~\bibnamefont {Stoll}},
  \bibinfo {author} {\bibfnamefont {A.~S.}\ \bibnamefont {Semisalova}},
  \bibinfo {author} {\bibfnamefont {S.}~\bibnamefont {Finizio}}, \bibinfo
  {author} {\bibfnamefont {A.}~\bibnamefont {Gangwar}}, \bibinfo {author}
  {\bibfnamefont {M.}~\bibnamefont {Weigand}}, \bibinfo {author} {\bibfnamefont
  {M.}~\bibnamefont {Noske}}, \bibinfo {author} {\bibfnamefont
  {M.}~\bibnamefont {F{\"a}hnle}}, \bibinfo {author} {\bibfnamefont
  {I.}~\bibnamefont {Bykova}}, \bibinfo {author} {\bibfnamefont
  {J.}~\bibnamefont {Gr{\"a}fe}}, \bibinfo {author} {\bibfnamefont {D.~A.}\
  \bibnamefont {Bozhko}}, \bibinfo {author} {\bibfnamefont {H.~Y.}\
  \bibnamefont {{Musiienko-Shmarova}}}, \bibinfo {author} {\bibfnamefont
  {V.}~\bibnamefont {Tiberkevich}}, \bibinfo {author} {\bibfnamefont {A.~N.}\
  \bibnamefont {Slavin}}, \bibinfo {author} {\bibfnamefont {C.~H.}\
  \bibnamefont {Back}}, \bibinfo {author} {\bibfnamefont {J.}~\bibnamefont
  {Raabe}}, \bibinfo {author} {\bibfnamefont {G.}~\bibnamefont {Sch{\"u}tz}},\
  and\ \bibinfo {author} {\bibfnamefont {S.}~\bibnamefont {Wintz}},\ }\bibfield
   {title} {\bibinfo {title} {Coherent {{Excitation}} of {{Heterosymmetric Spin
  Waves}} with {{Ultrashort Wavelengths}}},\ }\href@noop {} {\bibfield
  {journal} {\bibinfo  {journal} {PHYSICAL REVIEW LETTERS}\ } (\bibinfo {year}
  {2019})}\BibitemShut {NoStop}%
\bibitem [{\citenamefont {Onsager}(1931)}]{onsager_reciprocal_1931}%
  \BibitemOpen
  \bibfield  {author} {\bibinfo {author} {\bibfnamefont {L.}~\bibnamefont
  {Onsager}},\ }\bibfield  {title} {\bibinfo {title} {Reciprocal {Relations} in
  {Irreversible} {Processes}. {II}.},\ }\href
  {https://doi.org/10.1103/PhysRev.38.2265} {\bibfield  {journal} {\bibinfo
  {journal} {Physical Review}\ }\textbf {\bibinfo {volume} {38}},\ \bibinfo
  {pages} {2265} (\bibinfo {year} {1931})},\ \bibinfo {note} {copyright (C)
  2009 The American Physical Society; Please report any problems to
  prola@aps.org}\BibitemShut {NoStop}%
\bibitem [{\citenamefont {Kubo}(1966)}]{kubo_fluctuation-dissipation_1966}%
  \BibitemOpen
  \bibfield  {author} {\bibinfo {author} {\bibfnamefont {R.}~\bibnamefont
  {Kubo}},\ }\bibfield  {title} {\bibinfo {title} {The fluctuation-dissipation
  theorem},\ }\href {http://www.iop.org/EJ/abstract/0034-4885/29/1/306}
  {\bibfield  {journal} {\bibinfo  {journal} {Reports on Progress in Physics}\
  }\textbf {\bibinfo {volume} {29}},\ \bibinfo {pages} {255} (\bibinfo {year}
  {1966})}\BibitemShut {NoStop}%
\bibitem [{\citenamefont {Brown}(1963)}]{brown_thermal_1963}%
  \BibitemOpen
  \bibfield  {author} {\bibinfo {author} {\bibfnamefont {W.~F.}\ \bibnamefont
  {Brown}},\ }\bibfield  {title} {\bibinfo {title} {Thermal {Fluctuations} of a
  {Single}-{Domain} {Particle}},\ }\href
  {https://doi.org/10.1103/PhysRev.130.1677} {\bibfield  {journal} {\bibinfo
  {journal} {Physical Review}\ }\textbf {\bibinfo {volume} {130}},\ \bibinfo
  {pages} {1677} (\bibinfo {year} {1963})}\BibitemShut {NoStop}%
\bibitem [{\citenamefont {Zhang}\ and\ \citenamefont
  {Zhang}(2009)}]{zhangGeneralizationLandauLifshitzGilbertEquation2009}%
  \BibitemOpen
  \bibfield  {author} {\bibinfo {author} {\bibfnamefont {S.}~\bibnamefont
  {Zhang}}\ and\ \bibinfo {author} {\bibfnamefont {S.~S.-L.}\ \bibnamefont
  {Zhang}},\ }\bibfield  {title} {\bibinfo {title} {Generalization of the
  {{Landau-Lifshitz-Gilbert Equation}} for {{Conducting Ferromagnets}}},\
  }\href {https://doi.org/10.1103/PhysRevLett.102.086601} {\bibfield  {journal}
  {\bibinfo  {journal} {Physical Review Letters}\ }\textbf {\bibinfo {volume}
  {102}},\ \bibinfo {pages} {086601} (\bibinfo {year} {2009})}\BibitemShut
  {NoStop}%
\bibitem [{\citenamefont {Tserkovnyak}\ \emph {et~al.}(2009)\citenamefont
  {Tserkovnyak}, \citenamefont {Hankiewicz},\ and\ \citenamefont
  {Vignale}}]{tserkovnyakTransverseSpinDiffusion2009}%
  \BibitemOpen
  \bibfield  {author} {\bibinfo {author} {\bibfnamefont {Y.}~\bibnamefont
  {Tserkovnyak}}, \bibinfo {author} {\bibfnamefont {E.~M.}\ \bibnamefont
  {Hankiewicz}},\ and\ \bibinfo {author} {\bibfnamefont {G.}~\bibnamefont
  {Vignale}},\ }\bibfield  {title} {\bibinfo {title} {Transverse spin diffusion
  in ferromagnets},\ }\href {https://doi.org/10.1103/PhysRevB.79.094415}
  {\bibfield  {journal} {\bibinfo  {journal} {Physical Review B}\ }\textbf
  {\bibinfo {volume} {79}},\ \bibinfo {pages} {094415} (\bibinfo {year}
  {2009})}\BibitemShut {NoStop}%
\bibitem [{\citenamefont {Yuan}\ \emph {et~al.}(2019)\citenamefont {Yuan},
  \citenamefont {Liu}, \citenamefont {Xia}, \citenamefont {Yuan},\ and\
  \citenamefont {Wang}}]{yuanProperDissipativeTorques2019}%
  \BibitemOpen
  \bibfield  {author} {\bibinfo {author} {\bibfnamefont {H.~Y.}\ \bibnamefont
  {Yuan}}, \bibinfo {author} {\bibfnamefont {Q.}~\bibnamefont {Liu}}, \bibinfo
  {author} {\bibfnamefont {K.}~\bibnamefont {Xia}}, \bibinfo {author}
  {\bibfnamefont {Z.}~\bibnamefont {Yuan}},\ and\ \bibinfo {author}
  {\bibfnamefont {X.~R.}\ \bibnamefont {Wang}},\ }\bibfield  {title} {\bibinfo
  {title} {Proper dissipative torques in antiferromagnetic dynamics},\ }\href
  {https://doi.org/10.1209/0295-5075/126/67006} {\bibfield  {journal} {\bibinfo
   {journal} {EPL (Europhysics Letters)}\ }\textbf {\bibinfo {volume} {126}},\
  \bibinfo {pages} {67006} (\bibinfo {year} {2019})}\BibitemShut {NoStop}%
\end{thebibliography}%

\newpage
\appendix

\section{Relation between magnetic damping and polarization}
Consider a discrete system with $N$ sites.
The energy function of linearized spin wave is quadratic and can be written as 
\begin{equation}
    H=\frac{1}{2}\hPsi^\dagger \begin{pmatrix}
        A & B \\
        B^* & A^*
    \end{pmatrix}\hPsi=\frac{1}{2}\hPsi^\dagger h \hPsi,
\end{equation}
where $\hPsi=(\hpsi_1,\hpsi_2,\cdots,\hpsi_N,\hpsi^*_1, \hpsi^*_2,\cdots,\hpsi^*_N)^\text{T}$ and $\hpsi_j=\delta m_{j}^{(\theta)}+i\delta m_j^{(\phi)}$ is the local magnon creation operator. It satisfies the commutation relation $\qty[\hpsi^*_j,\hpsi_k]=\delta_{j,k}$ and other commutators are all zero. $A$ is the self-energy and hopping term, and $B$ is the squeezing term.
The Heisenberg equation of motion is an eigenvalue problem,
\begin{equation}
    -i\frac{d}{dt}\hPsi=\Sigma_z h\hPsi,
\end{equation}
where $\Sigma_z=\sigma_z \otimes I_{N \times N}$ and $\sigma_z$ and $I$ are the Pauli z matrix and identity matrix, respectively.

We can get the magnon frequencies by solving the equation of motion using Bogoliubov transformation,
\begin{equation}
    U^{-1}\Sigma_z h U =\begin{pmatrix}
        \Lambda &0 \\
        0 & -\Lambda 
    \end{pmatrix},
\end{equation}
with the eigenvectors
\begin{equation}
    \hPhi=U^{-1}\hPsi.
    \label{eq:Bogoliubov}
\end{equation}
$\Lambda$ is semi-positive defined diagonal matrix $\Lambda=\text{diag}(\lambda_1,\lambda_2,\cdots,\lambda_N)$.
We can label eigenvalues in the right below block in sequence. Its index $j$ is larger than $N$ and we have $\lambda_j=-\lambda_{j-N}$.  
The eigenvalue $\lambda_i$, $1\le i \le N$, is nonnegative to ensure the stability of ground state.
    
The matrix $U$ can be written as 
\begin{equation}
    U=\begin{pmatrix}
        M & N \\
        N^* & M^*
    \end{pmatrix}.
\end{equation}
To ensure the bosonic commutation relation, it is an element of U(N, N), 
\begin{equation}
    U^\dagger \Sigma_z U=\Sigma_z.
\end{equation}
It is easy to verify its inverse,
\begin{equation}
    U^{-1}=\Sigma_z U^\dagger \Sigma_z =\begin{pmatrix}
        M^\dagger & -N^\text{T} \\
        -N^\dagger & M^\text{T}
    \end{pmatrix}.
\end{equation}

Suppose that the eigenvector of $i$th eigenvalues $\lambda_i$ is $\ket{\Phi_i} e^{i\lambda_i t}$. It satisfies the eigenvalue equation,
\begin{equation}
    \Sigma_z h \ket{\Phi_i}=\lambda_i\ket{\Phi_i}.
\end{equation}
Note that the effective Hamiltonian $\Sigma_z h$ is non-Hermitian, the orthonormal relation is different from the Hermitian case,
\begin{equation}
    \braket{\Phi_i\Sigma_z}{\Phi_j}=\qty(\Sigma_z)_{i,j}.
\end{equation}

Now consider the effect of Gilbert damping with damping parameter $\alpha$. 
$\alpha=\text{diag}(\alpha_1,\alpha_2\cdots,\alpha_N,\alpha_1,\alpha_2\cdots,\alpha_N)$ is a diagonal matrix. Different labels of $\alpha$ indicate that the Gilbert damping could be different for different sites.
The damping term adds a damping matrix to the equation of motion
\begin{equation}
    -i\qty(1-i\alpha \Sigma_z)\frac{d}{dt}\hPsi=\Sigma_z h \hPsi.
\end{equation}
It is still a eigenvalue problem, 
\begin{equation}
    -i\frac{d}{dt}\hPsi=(1-i\alpha\Sigma_z)^{-1}\Sigma_z h\hPsi= \qty(\Sigma_z h+i\alpha h +O(\alpha^2))\hPsi.
\end{equation}
We don't need to solve the whole problem. Treat $\alpha$ as a small number, we are interested in the first order correction of the eigenvalues $\lambda_i$. The zeroth order correction is $\lambda_i$ itself $\lambda_i^{(0)}=\lambda_i$. The first order correction is given by perturbation theory, for $i \leq N$,
\begin{equation}
    \lambda_i^{(1)}= i \bra{\Phi_i \Sigma_z }\ket{\alpha h|\Phi_i}= i \lambda_i^{(0)}\braket{\Phi_i}{\alpha|\Phi_i}.
\end{equation}
For $N<i\leq 2N$, the first order correction is similar,
\begin{equation}
    \lambda_i^{(1)}=-i \lambda_i^{(0)}\braket{\Phi_i}{\alpha|\Phi_i}=i|\lambda_i^{(0)}|\braket{\Phi_i}{\alpha|\Phi_i}.
\end{equation}
We just need to focus on the first N states, because the other N states have the same decay rates with their conjugation partners.
The eigenfrequency with damping considered is then $\omega_i=\lambda_i + i\lambda_i \braket{\Phi_i}{\alpha|\Phi_i}$ for $1\leq i\leq N$. 
The ratio between the imaginary part and the real part of $\omega_i$ gives the $\beta$ parameter,
\begin{equation}
    \beta_i=\frac{\Im{\omega_i}}{\Re{\omega_i}}=\braket{\Phi_i}{\alpha|\Phi_i}.
\end{equation}
If the normalization of eigenvectors isn't strictly guaranteed, which is always the case in practice, we need to divide the metric norm of vector when getting the first order correction,
\begin{equation}
    \lambda_i^{(1)}=i \frac{\braket{\Phi_i}{\alpha|\Phi_i}}{\braket{\Phi_i\Sigma_z}{\Phi_i}}\lambda_i^{(0)}.
\end{equation}
Thus, the $\beta$ parameter is
\begin{equation}
    \beta_i= \frac{\braket{\Phi_i}{\alpha|\Phi_i}}{\braket{\Phi_i\Sigma_z}{\Phi_i}}.
\end{equation}

Now parametrize the eigenvector $\ket{\Phi_i}$ as $\ket{\Phi_i}=(M_{1,i},M_{2,i},\cdots,M_{N,i},N^*_{1,i},N^*_{2,i},\cdots,N^*_{N,i})^\text{T}$. Then, the metric norm is 
\begin{equation}
    \braket{\Phi_i\Sigma_z}{\Phi_i}=\sum_{j=1}^N |M_{j,i}|^2-|N^*_{j,i}|^2.
\end{equation}
The $\beta$ parameter can be written as 
\begin{equation}
    \beta_i= \frac{ \sum_{j=1}^N\alpha_j\qty(|M_{j,i}|^2+|N^*_{j,i}|^2)}{ \sum_{j=1}^N|M_{j,i}|^2-|N^*_{j,i}|^2}.
    \label{eq:beta_sum}
\end{equation}

The time evolution of the $j$th sites can be extracted from the Bogoliubov transformation \Eq{eq:Bogoliubov}. If the system is in the coherent state of $k$th mode at $t=0$, $\hPhi_k\ket{\xi}=\xi\ket{\xi}$, the time evolution of the $j$th site is 
\begin{equation}
    \expval{\psi_j(t)}=M_{j,k}\xi^* e^{i\lambda_k t}+N_{j,k}\xi e^{-i\lambda_k t}.
\end{equation}
We may assume $\xi$ to be real, since its argument can be absorbed into the $e^{i\lambda_k t}$ term.
The real part and imaginary parts of $\expval{\psi_j}$ correspond to two orthogonal components of the local magnon field. The trajectory of $\expval{\psi_j(t)}$ is an ellipse on complex plane with its center on the origin. We label the lengths of semi-major axis and semi-minor axis as $a_j$ and $b_j$ respectively. 
The trajectory can be parametrized as 
\begin{equation}
    \expval{\psi_j(t)}=e^{i\theta_j}\qty[a_j\cos(\lambda_k t+\phi_j)+ib_j\sin(\lambda_k t+\phi_j)],
\end{equation}
\begin{subequations}
    \begin{align}
        a_j=\xi\qty(|M_{j,k}|+|N_{j,k}|),\\
        b_j=\xi\qty(|M_{j,k}|-|N_{j,k}|),\\
        \theta_j=\frac{1}{2}\arg(M_{j,k} N_{j,k})\\
        \phi_j=\frac{1}{2}\arg(M_{j,k}N_{j,k}^*).
    \end{align}
\end{subequations}
We can verify that $a_j^2+b_j^2=2\xi^2\qty(|M_{j,k}|^2+|N_{j,k}|^2)$ and $a_j b_j=\xi^2|M_{j,k}|^2-|N_{j,k}|^2.$ 
Note that for $|M_{j,k}|>|N_{j,k}|$, we have $a_j>b_j>0$ and the precessing is right-handed. For $|M_{j,k}|<|N_{j,k}|$, we have $a_j>0>b_j$ and the precessing is left-handed.

If we define a signed area of the ellipse, $S_j=\pi a_j b_j$ based on the precession direction of local magnon. For right-handed precession, $a_j>b_j>0$, we have positive area $S_j>0$. For left-handed precession, $a_j>0>b_j$, we have negative area $S_j<0$. Then, we can calculate the $\beta$ parameter based on the ellipse trajectory of each site,
\begin{equation}
    \beta= \frac{\sum_{j=1}^N \alpha_j\qty(a_j^2+b_j^2)}{\sum_{j=1}^N 2a_j b_j}.
\end{equation}

Recall that $\eta=\frac{b}{a}=e^{-2r}$ in the main text. We can rewrite the numerator with $\eta$,
\begin{equation}
    a_j^2+b_j^2=a_j b_j(\eta+\eta^{-1})=2a_j b_j \cosh(2r_j).
\end{equation}
Note that, for left-handed precession sites, $\cosh(2r_j)$ is negative and the whole numerator $a_j b_j\cosh(2r)$ is still positive. 
Finally, we get the $\beta$ parameter,
\begin{equation}
    \beta=\frac{\sum_{j=1}^N \pi \alpha_j a_j b_j \cosh(2r_j)}{\sum_{j=1}^N \pi a_j b_j}=\frac{\sum_{j=1}^N S_j \alpha_j \cosh(2r_j)}{\sum_{j=1}^N S_j}.
\end{equation}
For a single site, we don't need the summation. The weight $S_j$ is then cancel out, $\beta=\alpha \cosh(2r)$.
In continuous limit, the summation becomes an integral,
\begin{equation}
    \beta=\frac{\int \dd^3{\bx}~S_\bx \alpha_\bx \cosh(2r_\bx)}{\int \dd^3{\bx}~S_\bx}..
\end{equation}
We get the \Eq{eqn:dpc} in the main text.



\section{Phase delay induced energy flow}
The Gilbert damping is local and related to the polarization of the local spin, \ie $\alpha \cosh(2r)$. While the system shares the same damping rate in eigenstate. The polarization of different place is different, which means that the energy loss rate by Gilbert damping is different, so there should be another mechanism to balance this difference.

Consider the two-macrospin model as an example. The free energy is 
\begin{equation}
    F = - \frac{K}{2} \sum_{i=1,2}(\mb_i\cdot\hbz)^2 
    + \frac{K'}{2} (\mb_2\cdot\hby)^2 - J \mb_1 \cdot \mb_2.
\end{equation}
The effective fields for lattice 1 and 2 are
\begin{equation}
    \begin{aligned}
        H_{1}&=K(\mb_1\cdot \hbz)\hbz+J\mb_2\\
        H_{2}&=K(\mb_2\cdot \hbz)\hbz-K'(\mb_2\cdot \hy)\hy+J\mb_1.
    \end{aligned}
\end{equation}
The ground state is a ferromagnet $\mb_{10}=\mb_{20}=\hbz$. Consider the spin wave excitation around ground state $\delta \mb_i=\mb_i-\mb_{i0}\approx(m_i^x,m_i^y,0)$ and $|\delta \mb_i|\ll 1$. With the help of complex combination $\psi_j=m_j^x+im_j^y$, we can write the linearized LLG equation, 
\begin{equation}
    \begin{aligned}
        -i(1-i\alpha_1)\dot \psi_1&=(K+J)\psi_1-J\psi_2,\\
        -i(1-i\alpha_2)\dot \psi_2&=(K+J)\psi_2+\frac{K'}{2}(\psi_2-\psi_2^*)-J\psi_1.
    \end{aligned}
\end{equation}
The solutions of these equations give ellipse trajectories $\psi_j=m_j^+ e^{i\omega t}+m_j^- e^{-i\omega t}$ with the lengths of two principal axes $a_j$ and $b_j$.

When the damping is neglected, the tangent direction of the trajectory is $\dot \mb_{j}$ and $\mb_{j0}\cdot(\dot \mb_j \times\tau)$ gives the area decrease due to the torque $\tau$ in time $t\sim t+dt$. The linearized form under complex combination is $\Im(\dot\psi_j \tau^*)$. The integral average of the ratio of $\Im(\dot\psi_j \tau^*)$ and its trajectory area over one period give the total relative decrease 
\begin{equation}
    \cA_j(\tau)=\frac{\omega}{2\pi}\int_0^{2\pi/\omega} \frac{\Im(\dot\psi_j \tau^*)}{a_jb_j\omega}dt.
\end{equation}

If damping is considered, the eigenfrequency gets an imaginary part $\Re{\omega}=\omega_0$ and $\Im{\omega}=\omega'$. The ellipse shrinks and the lengths of two axes decrease exponentially $a_j(t)=a_j(0)e^{-\omega't}$.
The time derivative of $\psi_j$ compensated with $\omega'\psi_j$ is then on the tangent direction of the ellipse,
\begin{equation}
    \cT_j=\dot\psi_j +\omega'\psi_j.
\end{equation}
The area decrease is then $\Im(\cT_j \tau^*)$.
Note that $\Im(\cT_j\psi_j^*)=\omega_0 a_j b_j e^{-2\omega' t}$ is proportion to the corresponding ellipse area at time t. The relative area decrease in one period is the integral average of the ratio between them,
\begin{equation}
    \cA_j(\tau)=\frac{\omega_0}{2\pi}\int_0^{2\pi/\omega_0} \frac{\Im(\cT_j \tau^*)}{\Im(\cT_j \psi_j^*)}dt.
\end{equation}
We could examine the terms in LLG equation one by one. 
The time derivative of $\psi_j$ gives the total damping
\begin{equation}
    \cA_j(\dot{\psi}_j)=\frac{\omega_0}{2\pi}\int_0^{2\pi/\omega_0} \frac{\Im(\cT_j \dot\psi_j^*)}{\Im(\cT_j \psi_j^*)}dt=-\omega'.
\end{equation}
The Gilbert damping term $i\alpha_j \dot\psi_j$ gives the area loss due to Gilbert damping, \ie the trajectory polarizations,
\begin{equation}
    \cA_j(i\alpha_j \dot\psi_j)=
    -\alpha_j\omega_0\frac{a_j^2+b_j^2}{2a_jb_j}=-\alpha_j\omega_0\cosh(2r_j).
\end{equation}
The torque from local anisotropy is a linear composition of $\psi_j$ and $\psi_j^*$, $\tau_A=i u_1 \psi_j+iu_2\psi_j^*$, with $u_1 \in \mathbb{R}$ and $u_2\in \mathbb{C}$. Its contribution to the area loss is always zero
\begin{equation}
    \cA_j(\tau_A)=0.
\end{equation}
The exchange torque from two lattices with collinear ground state is $\tau_E=iJ(\psi_j-\psi_{l})$. Its contribution is related to the phase delay of two lattices
\begin{equation}
        \cA_{jl}(\tau_\text{E})=\frac{J\Im{m_j^+ m_l^{+*}-m_j^- m_l^{-*}}}{a_j b_j}
\end{equation} 
$\cA(\tau)$ is a linear function, $\cA(c_1\tau_1+c_2\tau_2)=c_1\cA(\tau_1)+c_2\cA(\tau_2)$ for $c_1, c_2\in \mathbb{R}$. 
Apply it to LLG equations of two macrospin model, we find that 
\begin{equation}
    \cA_1(\dot\psi_1)=\cA_1(i\alpha_1 \dot\psi_1)+\cA_{12}(\tau_\text{E}).
\end{equation}
It means that the total area loss is equal to the sum of loss due to Gilbert damping and the transfer between two lattices.
We verify this result by examining it on the numerical solution of two macrospin model.

If we compensated the exponential decay part of the eigenvector, we could neglect the denominator since $\Im(e^{2\omega' t}\cT_j \psi_j^*)=\omega_0a_jb_j$ is a constant. Using the compensated eigenstate $\tilde{\psi_j}=e^{\omega' t}\psi_j$, we have $\cT_j=e^{-\omega' t}\dot{\tilde{\psi}}_j$ we could define the absolute area loss, 
\begin{equation}
    \tilde{\cA}_j(\tau)=\frac{\omega_0}{2\pi}\int_0^{2\pi/\omega_0} \Im(\dot{\tilde{\psi}}_j \tau^*)dt.
\end{equation}
The transfer between two lattices could be written as 
\begin{equation}
    \tilde{\cA}_{jl}(\tau_\text{E})=\expval{\Im{iJ\dot{\tilde{\psi}}_j(\tilde{\psi}_j-\tilde{\psi}_l)^*}}=\expval{\Im{-iJ\dot{\tilde{\psi}}_j\tilde{\psi}_l^*}}.
\end{equation}
Then we have 
\begin{equation}
    \omega' \omega_0 a_j b_j=\expval{\alpha_j|\dot{\tilde{\psi}}_j|^2}+\expval{\Im{iJ\dot{\tilde{\psi}}_j\tilde{\psi}_l^*}}.
\end{equation}
Its physical meaning is the same as above.  

Now focus on the Heisenberg exchange transfer term, and suppose $\mb$ represents the compensated magnetic vector, we have
\begin{equation}
    \begin{aligned}
        \tilde{\cA}_{12}(\tau_\text{E})&=-\frac{\omega_0J}{2\pi}\int_0^{2\pi/\omega_0}\hbz\cdot(\dot{\mb}_1\times(\mb_1\times \mb_2)) dt\\
        &=-\frac{\omega_0J}{2\pi}\int_0^{2\pi/\omega_0} \dot{\mb}_1\cdot \mb_2dt\\
        &=-J\Im{m_1^+ {m_2^+}^* - m_1^- {m_2^-}^*}
    \end{aligned}
\end{equation}
It gives the flow from lattice 1 to lattice 2, which balance the mismatch of area loss from Gilbert damping due to space varies polarization.
\end{document}